\documentclass[12pt]{article}
\pdfoutput=1
\usepackage{jcappub}

\usepackage{amsmath,amsfonts,amssymb,mathrsfs,graphicx,color,longtable}
\usepackage{hyperref}
\usepackage{graphicx}
\usepackage{amsfonts,amsmath,amssymb,bm}
\usepackage{array}
\usepackage{color}
\usepackage{enumitem}
\usepackage[explicit]{titlesec}
\usepackage[utf8]{inputenc}
\usepackage[normalem]{ulem}

\hypersetup{
     colorlinks   = true,
     citecolor    = red,
     urlcolor     = blue,
     linkcolor    = blue
}

\usepackage{xcolor}
\usepackage{mhchem}

\newcommand{\dd}{\mathrm{d}}

\begin{document}
\title{Phantom scalar-tensor models and cosmological tensions}

\author[1,2,3]{Mario Ballardini,}
\author[4,5]{Angelo G. Ferrari,}
\author[3,5]{Fabio Finelli}

\affiliation[1]{Dipartimento di Fisica e Scienze della Terra, Universit\`a degli Studi di Ferrara, via Giuseppe Saragat 1, 44122 Ferrara, Italy}
\affiliation[2]{INFN, Sezione di Ferrara, via Giuseppe Saragat 1, 44122 Ferrara, Italy}
\affiliation[3]{INAF/OAS Bologna, via Piero Gobetti 101, 40129 Bologna, Italy}
\affiliation[4]{Dipartimento di Fisica e Astronomia, Universit\`a di Bologna, viale Berti Pichat 6/2, 40127 Bologna, Italy}
\affiliation[5]{INFN, Sezione di Bologna, viale C. Berti Pichat 6/2, 40127 Bologna, Italy}

\emailAdd{mario.ballardini@unife.it}
\emailAdd{angelo.ferrari3@unibo.it}
\emailAdd{fabio.finelli@inaf.it}

\abstract{
We study three different extended scalar-tensor theories of gravity by also allowing a negative sign for 
the kinetic term for the scalar field in the Jordan frame. Our scope is to understand how the observational 
constraints for these models cope with the volume of the parameter space in which the theory is healthy. 
Models with a negative kinetic term lead to decreasing effective gravitational constant with redshift and 
behave as an effective relativistic component with a negative energy density as opposite to their corresponding 
version with a standard kinetic term. As a consequence, we find that the extended branch with a negative sign 
for the kinetic term correspond in general to lower $H_0$ and $\sigma_8$ compared to $\Lambda$CDM. We find 
that in all the cases with a negative sign for the kinetic term studied here, cosmological observations 
constrain these models around GR and prefer a volume of the parameter space in which the theory is not healthy 
since the scalar field behave as a ghost also in the related Einstein frame.
We show that also in the phantom branch early modify gravity with a quartic coupling can substantially 
 reduce the $H_0$ tension fitting the combination of cosmic microwave background data from {\em Planck}, 
 baryon acoustic oscillations from BOSS and eBOSS, and Supernovae from the Pantheon sample with calibration 
 information by SH0ES.
}

\maketitle

\section{Introduction}
The $\Lambda$CDM model represents the current standard cosmological model providing an excellent fit to most of 
cosmological observations: measurements of luminosity distances of Type Ia Supernovae (SN Ia) 
\cite{SupernovaSearchTeam:1998fmf,SupernovaCosmologyProject:1998vns,Carr:2021lcj}, measurements of cosmic microwave 
background (CMB) anisotropies in temperature and polarization \cite{Planck:2018nkj}, measurements of the baryon 
acoustic oscillations (BAO) in galaxy and cluster distribution \cite{eBOSS:2020yzd,Veropalumbo:2013cua}, cosmic 
shear measurements of the CMB \cite{Sherwin:2016tyf,Wu:2019hek,Darwish:2020fwf} and of the galaxy distribution 
\cite{KiDS:2020suj,DES:2021wwk}, and the predicted abundance of light elements \cite{Cooke:2017cwo}. 
While the $\Lambda$CDM model provides an accurate description to most of cosmological observations, it relies on a 
number of assumptions and unknown ingredients such dark matter, dark energy, and a suitable mechanism to produce 
its initial condition.

In addition to the interest in testing at which extent the validity of the $\Lambda$CDM model holds with better 
and more data, the theoretical search for extended models 
\cite{Umilta:2015cta,Rossi:2019lgt,Poulin:2018cxd,Agrawal:2019lmo,Hart:2019dxi,Jedamzik:2020krr,Ballardini:2020iws,Braglia:2020auw,Antony:2022ert} has been 
fueled by the persisting cosmological tensions or rather intriguing inconsistencies between different measurements 
under the framework of the minimal $\Lambda$CDM model; see 
Refs.~\cite{Knox:2019rjx,DiValentino:2020zio,DiValentino:2020vvd,DiValentino:2021izs,Perivolaropoulos:2021jda,Schoneberg:2021qvd,Shah:2021onj,Abdalla:2022yfr} 
for reviews on the topic.

Among the many proposed models, there are still difficulties in finding a candidate able to completely {\rm solve} 
the discrepancy between the value of the Hubble parameter inferred in $\Lambda$CDM using CMB data from the {\em Planck} 
DR3, i.e. $H_0 = (67.36\pm 0.54)\,{\rm km}\, {\rm s}^{-1}\, {\rm Mpc}^{-1}$ at 68\% confidence level (CL) 
\cite{Planck:2018vyg}, with the measurement from the SH0ES team \cite{Riess:2021jrx} obtained with cosmic distance 
ladder calibration of SN Ia from the revised Pantheon+ compilation \cite{Carr:2021lcj}, i.e. 
$H_0 = (73.0\pm 1.0)\,{\rm km}\, {\rm s}^{-1}\, {\rm Mpc}^{-1}$ at 68\% CL, once all cosmological data are combined. 

It is even more difficult to reconcile the value of the Hubble parameter together with the persistent but less 
significant tension between {\em Planck} and galaxy shear experiments, quantified through the value of 
$S_8\equiv \sigma_8\sqrt{\Omega_{\rm m}/0.3}$, see \cite{Schoneberg:2021qvd}. 
Adopting a flat $\Lambda$CDM model, cosmic shear analysis of the fourth data release of the {\em Kilo-Degree Survey} 
(KiDS-1000) reported $S_8 = 0.759^{+0.024}_{-0.021}$ \cite{KiDS:2020suj} and $S_8 = 0.776\pm0.017$ from {\em Dark 
Energy Survey} (DES) Year 3 (Y3) combination of three large-scale structures (LSS) two-point correlation functions 
($3\times 2$ pt) \cite{DES:2021wwk}, while the value measured by {\em Planck} corresponds to $S_8 = 0.832\pm 0.013$ 
\cite{Planck:2018vyg}.
Indeed, while minimally and nonminimally coupled scalar field have been extensively studied as possible solutions 
to the Hubble tension, they usually lead to a higher value of the Hubble constant together with a larger growth of 
structures on small scales, i.e. a higher value of $\sigma_8$, see 
Refs.~\cite{Umilta:2015cta,Ballardini:2016cvy,Rossi:2019lgt,Hill:2020osr,Braglia:2020iik,Ivanov:2020ril,Rezazadeh:2022lsf}. 
However, they generally predict a value of $S_8$ compatible to the one obtained in $\Lambda$CDM avoiding to exacerbate 
the tension on the growth of structure amplitude since the larger $\sigma_8$ is compensated by a larger value of $H_0$ 
and a lower value of $\Omega_{\rm m}$ \cite{Smith:2020rxx,Braglia:2020auw,Ballardini:2021evv,Simon:2022adh}.

One possibility is to extend the dynamics of the scalar field to behave differently at early- and late-time in order to 
solve both tensions at the same time.
The possibility to have models with phenomenology in both the early and late universe has been tried in the context of 
modified gravity \cite{Braglia:2020auw,Ballardini:2021evv}, early dark sector \cite{McDonough:2021pdg}, and combining 
modified gravity or early dark energy to extended neutrino physics 
\cite{Ballardini:2020iws,Gomez-Valent:2022bku,Reeves:2022aoi}.

In this paper, we study modified gravity models with a nonminimally coupled scalar field with negative kinetic energy, 
so-called {\em phantom} field. Note that this is not strictly related to the the phantom dark energy models for which 
the dark energy (DE) equation of state can cross the {\em phantom divide} line $w_{\rm DE} = -1$. 
Moreover, scalar-tensor models can be realized with no necessity to introduce a ghost field \cite{Boisseau:2000pr} 
avoiding the problems in ghost phantom DE \cite{Caldwell:1999ew} to be plagued by classical and quantum instabilities 
\cite{Cline:2003gs}.
Such a the non-canonical kinetic energy term can occur in supergravity models \cite{Nilles:1983ge} and in
higher derivative theories of gravity \cite{Pollock:1988xe}.
We show how a nonminimally coupled scalar field with the negative sign of its kinetic term (phantom branch) behaves 
differently compared to the case with standard kinetic term (standard branch) and we derive the constraints on these 
models combining the information from {\em Planck} 2018 DR3 CMB temperature, polarization and lensing, together with 
a compilation of BAO measurements from the releases DR7 and DR12 of the {\em Baryon Oscillation Spectroscopic Survey} 
(BOSS) and Ly$\alpha$ measurements from the {\em extended Baryon Oscillation Spectroscopic Survey} (eBOSS), and 
uncalibrated SN Ia from the Pantheon sample.

The paper is organized as follows. 
After this introduction, we describe the implementation of the various basic quantities in the context of 
scalar-tensor theories in Section~\ref{sec:theory}. In Section~\ref{sec:results}, we describe the datasets and prior 
considered and we discuss our results for the three models studies: induced gravity, non-minimal coupling, and early 
modified gravity. In Section~\ref{sec:conclusions} we draw our conclusions.
In Appendix~\ref{sec:app_tab}, we collect the tables with the constraints on all the cosmological parameters obtained 
with our MCMC analysis. Background equations, linear perturbations, and initial conditions for background and 
cosmological fluctuations are collected in Appendices~\ref{sec:app_bkg}-\ref{sec:app_pert}-\ref{sec:app_IC}. 
In Appendix~\ref{sec:app_bao}, we present a comparison of the results by using CMB data plus different combinations 
of LSS measurements.

\section{Theory and cosmological background dynamics} \label{sec:theory}
We study the action for the scalar-tensor theory in Jordan frame \cite{Gannouji:2006jm} which is given by
\begin{equation} \label{eqn:action}
    {\cal S} = \int {\dd}^4x\sqrt{|g|} \left[\frac{F(\sigma)R}{2} - \frac{Z(\sigma)}{2}(\partial\sigma)^2
                -V(\sigma) + {\cal L}_m\right]
\end{equation}
where $|g|$ is the absolute value of the determinant of the metric $g_{\mu\nu}$, $\sigma$ is the scalar field, 
$F(\sigma)$ is the non-minimal coupling function, $R$ is the Ricci scalar, $V(\sigma)$ is the potential for $\sigma$, 
and ${\cal L}_m$ the Lagrangian density of matter minimally coupled to the metric (without introducing any direct 
coupling between the scalar field and the matter content we guarantee that the weak equivalence principle is exactly 
satisfied). The function $Z(\sigma)$ in front of the kinetic term can be set to $\pm 1$ by a redefinition of the 
scalar field.

In this model, the effective gravitational constant $G_{\rm eff}$ for the attraction between two test masses [having 
the same physical meaning as the Newton gravitational constant in general relativity (GR)] is given by
\begin{equation} 
    G_{\rm eff} = \frac{1}{8\pi F}\frac{ZF + 2F^2_\sigma}{ZF + \frac{3}{2}F^2_\sigma}
\end{equation}
on all scales for which the scalar field is effectively massless \cite{Boisseau:2000pr}, i.e. 
$V_\sigma \simeq 0$ and $V_{\sigma\sigma} \simeq 0$.

The current values of the time derivative and field derivative 
of coupling $F$ in these theories - assuming a homogeneous evolution of the scalar field for all the scales - are 
strongly constrained by Solar System tests of post-Newtonian parameters (for these quantities, we drop here the 
subscript 0)
\begin{align}
    &\gamma_{\rm PN} = 1 - \frac{F_\sigma^2}{Z F + 2 F^2_\sigma} \\
    &\beta_{\rm PN} = 1 + \frac{1}{4}\frac{F F_\sigma}{2 Z F + 3 F^2_\sigma}
    \frac{\dd \gamma_{\rm PN}}{\dd \sigma}
\end{align}
as well as the time variation of the effective cosmological constant. Current constraints 
\cite{Bertotti:2003rm,Clifton:2005xr,Muller:2007zzb,Belgacem:2018wtb} correspond to 
\begin{align}
    &\gamma_{\rm PN} - 1 = \left(2.1 \pm 2.3\right) \cdot 10^{-5} \\
    &\beta_{\rm PN} - 1 = \left(-4.1 \pm 7.8\right) \cdot 10^{-4} \\
    &\dot{G}/G = \left(7.1 \pm 7.6\right) \cdot 10^{-14}\, {\rm yr}^{-1} \,.
\end{align}
On cosmological scales, post-Newtonian parameters are weakly constrained from current cosmological data, see 
Ref.~\cite{Ballardini:2020iws}, with the perspective to reach the Solar System accuracy with the combination of future 
cosmological surveys \cite{Alonso:2016suf,Ballardini:2019tho,Ballardini:2021evv}.

There are essentially two stability conditions which impact on these scalar-tensor theories. The condition 
\begin{equation} \label{eqn:ghostfree1}
    G_{\rm eff} > 0
\end{equation} 
is one of the stability conditions of this theory meaning that the graviton is not a ghost. 
Moreover, we have the inequality 
\begin{equation} \label{eqn:ghostfree2}
    \frac{Z F}{F_\sigma^2} > -\frac{3}{2}
\end{equation} 
requiring the positivity of the kinetic energy of the scalar field in the Einstein frame \cite{Esposito-Farese:2000pbo}. 
Eqs.~\eqref{eqn:ghostfree1}-\eqref{eqn:ghostfree2} reduce to $Z F F_\sigma^{-2} > 0$ for $Z = +1$ or 
$-3/2 < Z F F_\sigma^{-2} < 0$ for $Z = -1$, and to $F > 0$.
This condition can be mapped  to a range of allowed parameter space for the parameters modelling $F(\sigma)$. However, 
we will consider a larger parameter space in the following analysis testing the models also for parameters violating the 
stability conditions in an agnostic way.

\section{Constraints and results} \label{sec:results}
In this section, we present our constraints on the cosmological parameters of the models studied. In particular, we 
study the nonminimally coupling $F=\xi\sigma^2$, i.e. {\em induced gravity} (IG) \cite{Cooper:1981byv,Wetterich:1987fk}, 
and $F=N_{\rm Pl}^2+\xi\sigma^2$ (hereafter NMC) both with a phantom scalar field, i.e. $Z = -1$; in both cases we 
consider $V(\sigma) = \lambda F^2(\sigma)/4$ which yields to an effectively massless dynamic \cite{Amendola:1999qq,Finelli:2007wb}. 
We study also the {\em early modified gravity} (EMG) model proposed in Ref.~\cite{Braglia:2020auw} extended to $Z = -1$. 
This last case is given by $F=M_{\rm Pl}^2+\xi\sigma^2$ and $V=\Lambda + \lambda \sigma^4/4$ with a negative amplitude 
$\lambda$ of the self-interaction term in order to produce the peculiar evolution of the scalar field damped into coherent 
oscillations within the phantom branch. We perform a Markov-chain Monte Carlo (MCMC) analysis using a modified version of 
the {\tt CLASSig} code \cite{Umilta:2015cta}, based on the Einstein-Boltzmann code 
{\tt CLASS}\footnote{\href{https://github.com/lesgourg/class\_public}{https://github.com/lesgourg/class\_public}} 
\cite{Lesgourgues:2011re,Blas:2011rf}, interfaced to the publicly sampling code 
{\tt MontePython}\footnote{\href{https://github.com/brinckmann/montepython\_public}{https://github.com/brinckmann/montepython\_public}} \cite{Audren:2012wb,Brinckmann:2018cvx}. The datasets used in this work include
\begin{itemize}
    \item P18 refers to the CMB temperature, polarization, and lensing from {\em Planck} DR3 
    \cite{Aghanim:2019ame,Aghanim:2018oex}.
    \item FS refers to the combination of pre-reconstructed full-shape monopole and quadrupole galaxy power spectra for 
    three different sky-cuts CMASS NGC, CMASS SGC and LOWZ NGC \cite{Gil-Marin:2015sqa} based on     the publicly available code 
    {\tt PyBird}\footnote{\href{https://github.com/pierrexyz/pybird/}{https://github.com/pierrexyz/pybird/}}\cite{DAmico:2020kxu}.
    \item BAO refers to the post-reconstruction measurements from BOSS DR12 \cite{BOSS:2016wmc},     low-$z$ BAO measurements 
    from SDSS DR7 6dF and MGS \cite{Beutler:2011hx,Ross:2014qpa}, Ly$\alpha$ BAO  measurements from eBOSS, and combination of 
    those \cite{deSainteAgathe:2019voe,Blomqvist:2019rah,Cuceu:2019for}.
    \item SN refers to the Pantheon catalogue of high-redshift supernovae, spanning the redshift range $0.01 < z < 2.3$ 
    \cite{Pan-STARRS1:2017jku}\footnote{\href{https://github.com/dscolnic/Pantheon}{https://github.com/dscolnic/Pantheon}}.
    \item Additional constraints of a Gaussian prior on the density of baryons (hereafter BBN) motivated from Big Bang 
    nucleosynthesis (BBN) constraints     corresponding to $\omega_{\rm b} = 0.02235 \pm 0.0005$ \cite{Cooke:2017cwo}, used in 
    combination to FS and     SN for the CMB-independent analysis.
    \item Additional constraints that include a Gaussian prior on the Hubble constant [hereafter $p(H_0)$], 
    $H_0 = (73.04\pm 1.04)\,{\rm km}\,{\rm s}^{-1}\,{\rm Mpc}^{-1}$ at 68\% CL, from Ref.~\cite{Riess:2021jrx}.
\end{itemize}
We vary 6 standard parameters, i.e. $\omega_{\rm b}$, $\omega_{\rm c}$, $H_0$, $\tau$, $\ln\left(10^{10}A_{\rm s}\right)$, 
$n_{\rm s}$, and the modified gravity parameters. We assume 2 massless neutrino with $N_{\rm ur} = 2.0328$, and a massive 
one with fixed minimum mass $m_\nu = 0.06\,{\rm eV}$. We fix the primordial \ce{^{4}He} mass fraction $Y_{\rm p}$ according 
to the prediction from {\tt PArthENoPE} \cite{Pisanti:2007hk,Consiglio:2017pot}, by taking into account the relation with 
the baryon fraction $\omega_{\rm b}$ and the varying gravitational constant which enters in the Friedman equation during 
nucleosynthesis.

Following the minimization method of Ref.~\cite{Schoneberg:2021qvd}, we report for each combination of datasets the $\Delta \chi^2$ values calculated with respect to the $\Lambda$CDM model where negative values correspond to a better fit of the dataset.

\subsection{Phantom induced gravity}
For IG (or equivalently extended Jordan-Brans-Dicke), with coupling $F(\sigma) = \xi\sigma^2$, we sample on the quantity 
$\zeta_{\rm IG} \equiv \ln\left(1 + 4\xi\right)$ 
which corresponds to a linear prior on the coupling to the parameter $\xi$ for $\xi \ll 1$. 
Here we impose the following boundary condition on the current value of the effective gravitational constant 
\begin{equation} \label{eqn:boundary}
    G_{\rm eff}(z=0) = G
\end{equation}
which fixes the final value of the scalar field.

\begin{figure}
\centering
\includegraphics[width=0.49\textwidth]{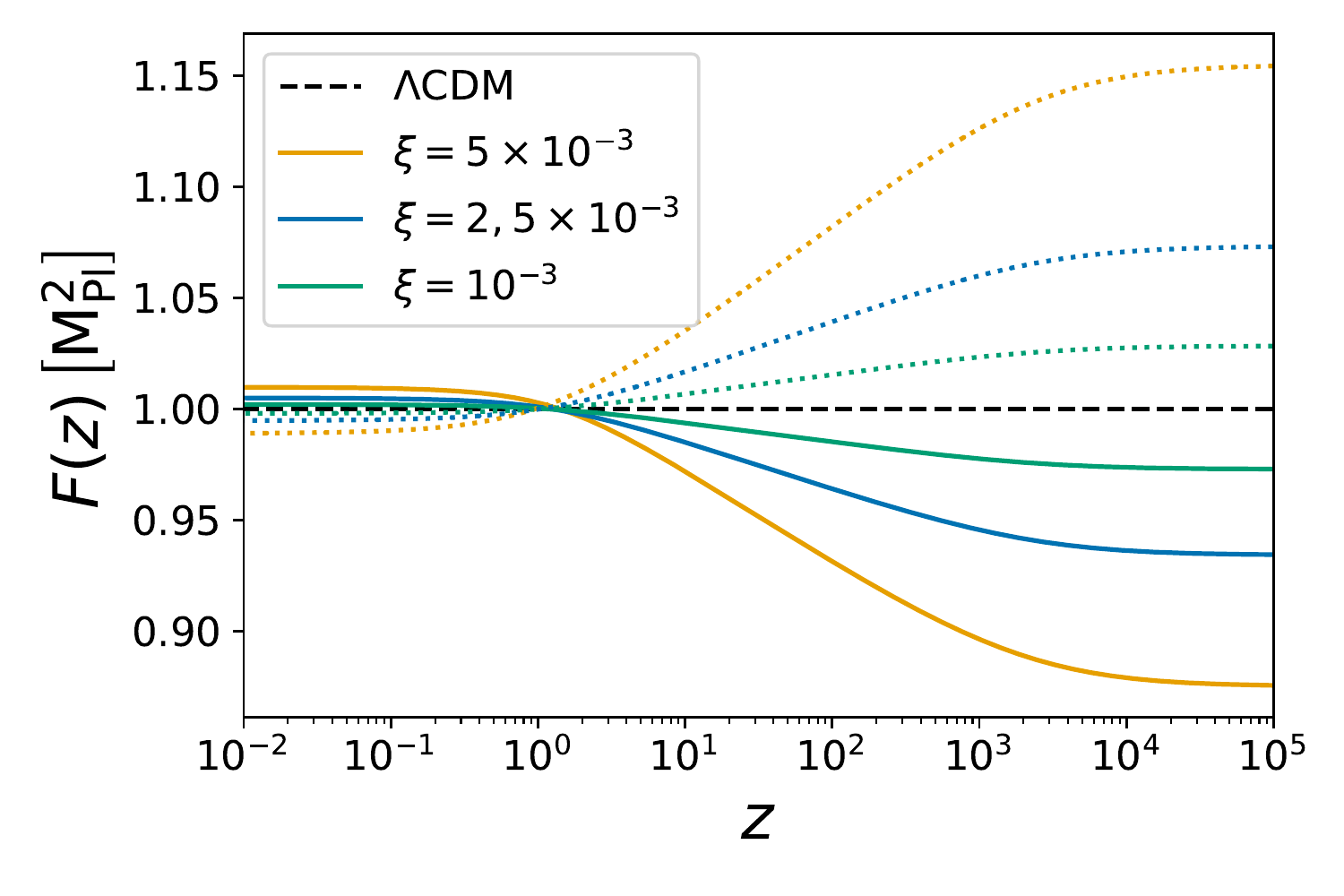}
\includegraphics[width=0.49\textwidth]{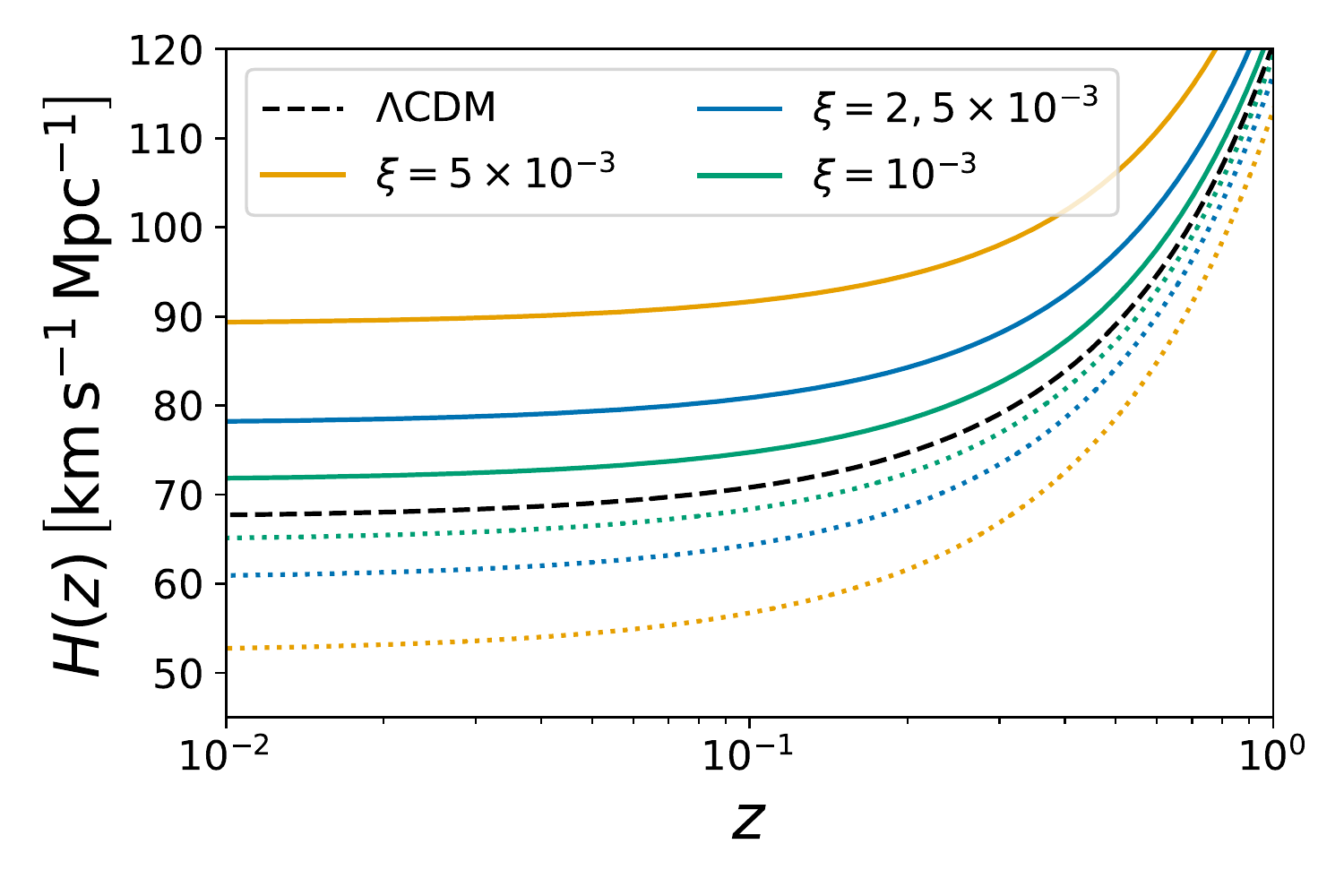}
\caption{Time evolution of the coupling to the Ricci scalar $F = \xi\sigma^2$ (left panel) and of the Hubble parameter 
(right panel) for different values of the coupling parameter $\xi$ in the standard branch (solid lines) and in the phantom 
one (dashed lines) for IG ($F = \xi\sigma^2,\ V = \lambda F^2/4$).}
\label{fig:FandH}
\end{figure}

\begin{figure}
\centering
\includegraphics[width=0.49\textwidth]{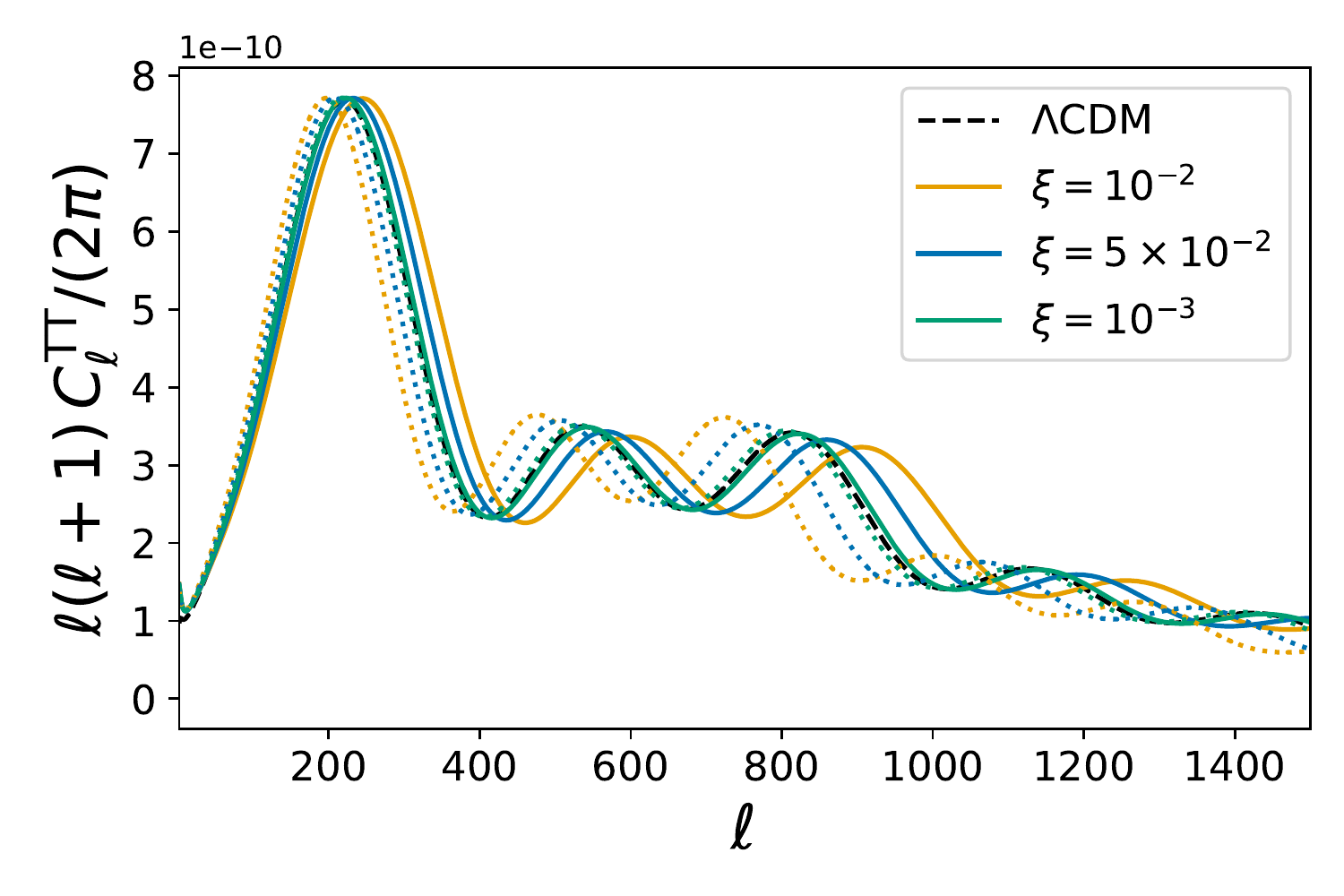}
\includegraphics[width=0.49\textwidth]{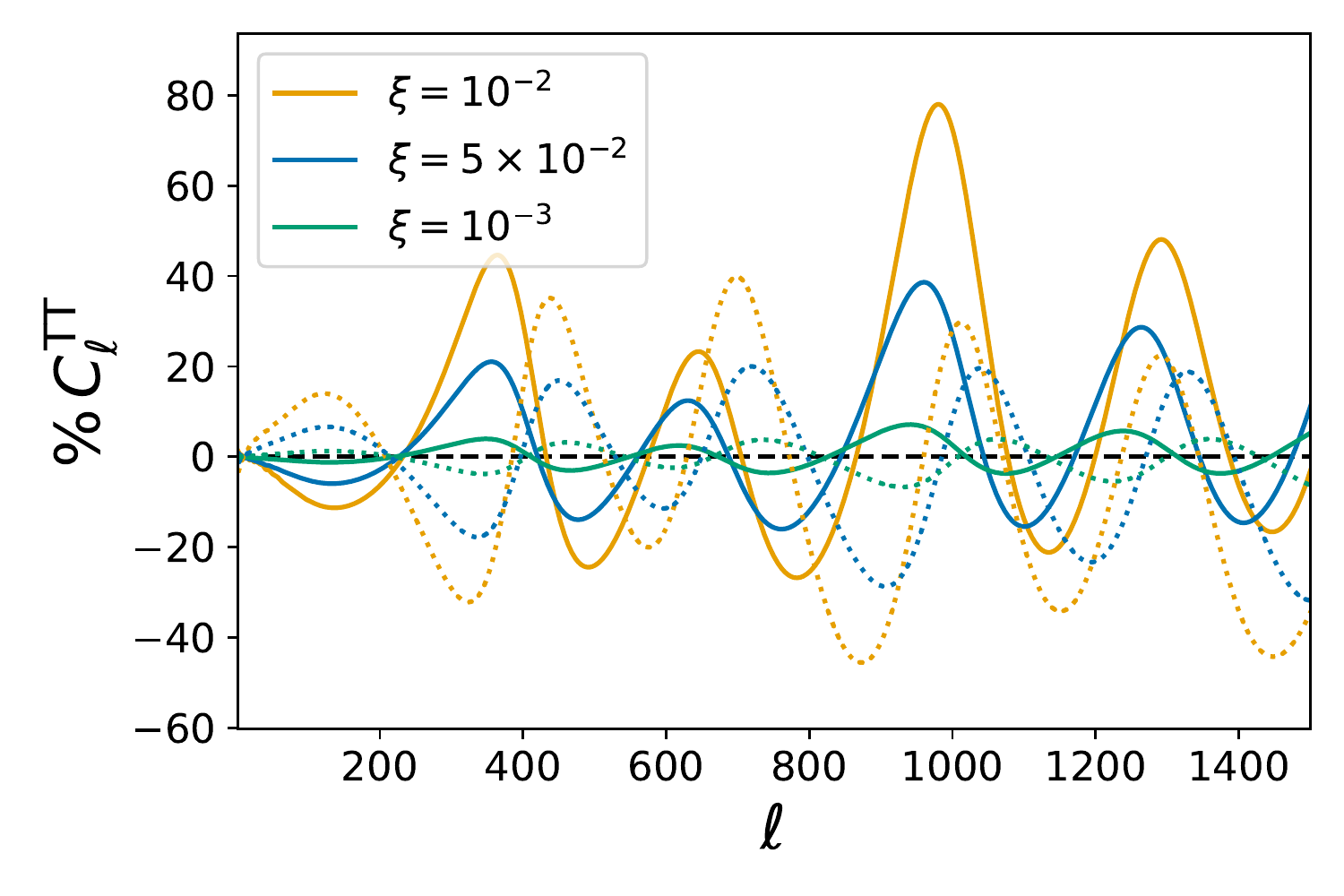}
\caption{CMB temperature anisotropies power spectrum (left panel) and relative differences with respect to the 
$\Lambda$CDM case (right panel) for different values of the coupling parameter $\xi$ in the standard branch (solid lines) 
and in the phantom one (dashed lines) for IG ($F = \xi\sigma^2,\ V = \lambda F^2/4$).}
\label{fig:CTT}
\end{figure}

\begin{figure}
\centering
\includegraphics[width=0.98\textwidth]{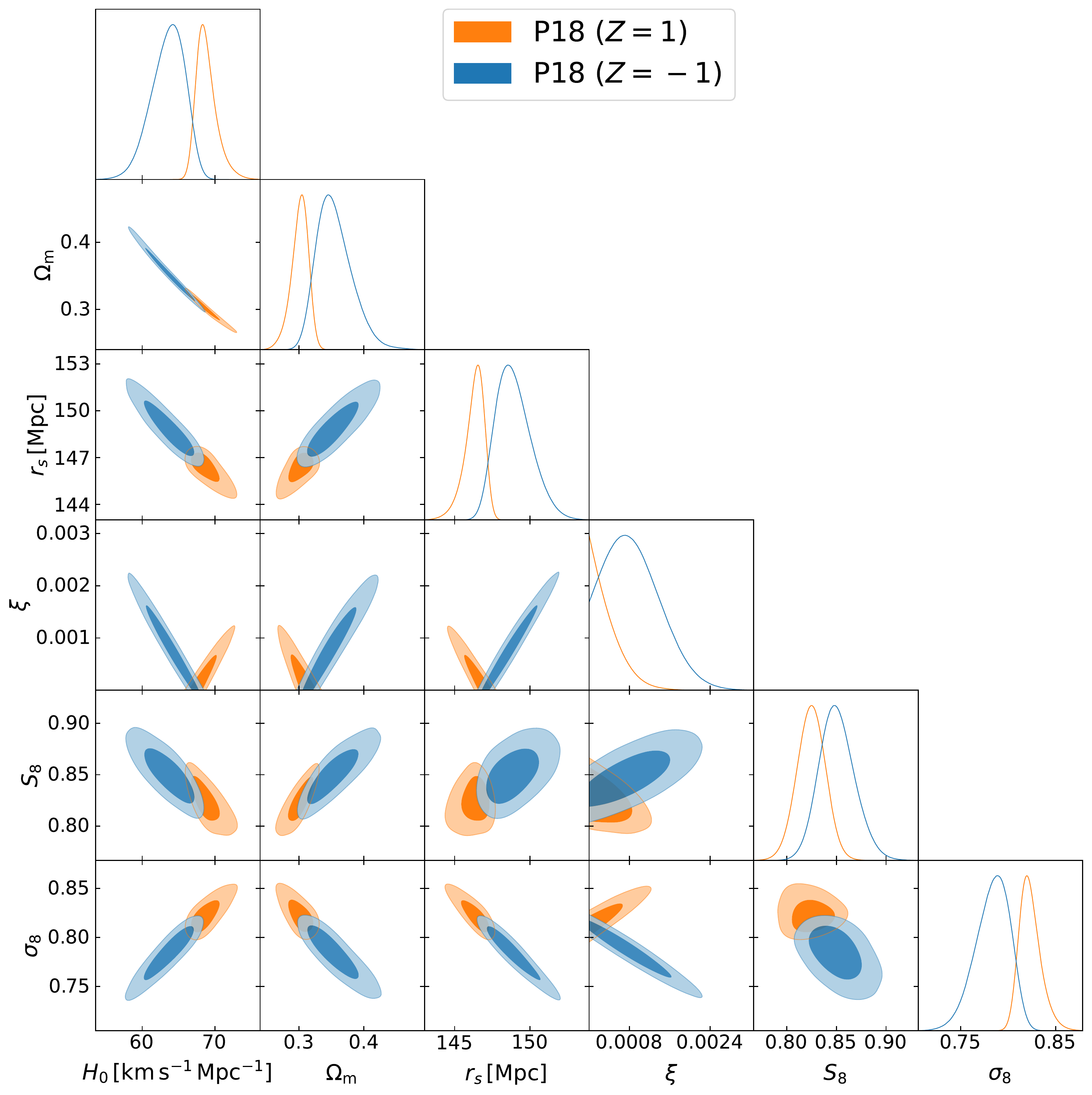}
\caption{Marginalized joint 68\% and 95\% CL regions 2D parameter space using the CMB alone data for IG 
($F = \xi\sigma^2,\ V = \lambda F^2/4$) with $Z = 1$ (orange) and for IG with $Z = -1$ (blue).}
\label{fig:ig_ph_comp}
\end{figure}

Scalar-tensor theories of gravity such as extended Jordan-Brans-Dicke models, lead to a modification of the Hubble 
parameter $H_0$ due to the time evolution of $\sigma$ and due to the redshift evolution of the gravitational strength.
Indeed, a variation of the strength of gravity can be connected to a change of the expansion rate of the universe as 
\begin{equation} \label{eqn:ig_H}
    \frac{H(\xi \ne 0)}{H(\xi = 0)} \approx \frac{M_{\rm Pl}^2}{F(\sigma)} \,.
\end{equation}
For a fixed matter content, reducing the Planck mass $F(\sigma) < M_{\rm Pl}^2$ with respect to the GR prediction 
increases the expansion rate at a given time and consequently reduces the comoving sound horizon at recombination
\begin{equation}
    r_s = \int_{z_*}^\infty \dd z' \frac{c_{\rm s}(z')}{H(z')}
\end{equation}
where $z_*$ is the redshift parameter at recombination and $c_{\rm s}$ is the speed of sound in the photon-baryon fluid.
We show in Fig.~\ref{fig:FandH} that the coupling function increases in the branch with standard kinetic term (solid lines) 
while decreasing in the phantom branch (dashed lines). 
This different behaviour is connected with a different late-time evolution of the Hubble parameter (when the scalar field 
starts to evolve driven by the non-relativistic matter) which is larger than the $\Lambda$CDM case in the standard branch 
and lower in the phantom branch.
This effect induces also a modification on the comoving angular diameter distance
\begin{equation}
    D_{\rm M}(z) =  \int_{0}^z \frac{\dd z'}{H(z')}
\end{equation}
and does not cancel out on the angular size of horizon at the last-scattering surface $\theta_*$ 
\begin{equation}
    \theta_* = \frac{r_s}{D_{\rm M}(z_*)}
\end{equation}
driving a shift on the acoustic peaks of the CMB connected to the evolution of the coupling $F$
\cite{Zumalacarregui:2020cjh,Ballardini:2020iws,SolaPeracaula:2020vpg}. 
In Fig.~\ref{fig:CTT}, we show the shift of the acoustic peaks of the CMB temperature anisotropies angular power spectrum 
imprinted by the evolving effective Planck mass. The peaks move to the right in the positive branch and to the left in the 
phantom one. Indeed, in order to compensate this shift (keeping nearly unchanged the value of the CDM density parameter) 
the two branches go in the direction of a larger or a smaller value of $\Omega_{\rm m}$ once the CMB data are included in 
the analysis, see Fig.~\ref{fig:ig_ph_comp}.
Therefore it is possible to break the degeneracy at background level between the scalar field $\sigma$ and the density 
parameters by combining early- and late-time probes.

\begin{figure}
\centering
\includegraphics[width=0.98\textwidth]{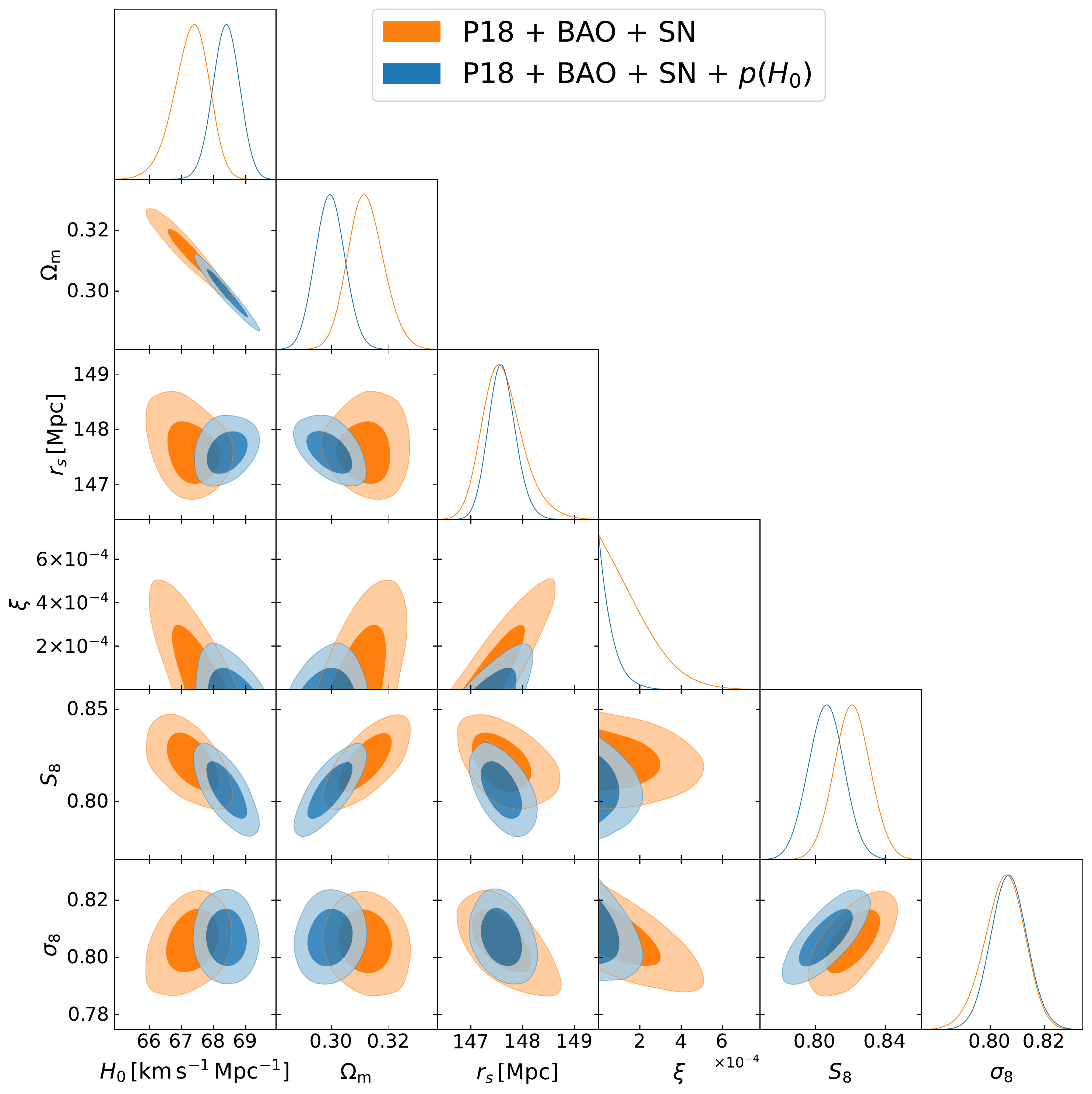}
\caption{Marginalized joint 68\% and 95\% CL regions 2D parameter space using the combination CMB+BAO+SN (orange) and 
CMB+BAO+SN+$p(H_0)$ (blue) for IG ($F = \xi\sigma^2,\ V = \lambda F^2/4$) in the phantom branch ($Z = -1$).}
\label{fig:ig_ph_H0}
\end{figure}
In Fig.~\ref{fig:ig_ph_comp}, we compare the CMB-only constraints for IG in the phantom branch to the standard case. We 
see that the degeneracy direction in the $\xi$-$H_0$ plane changes orientation going from one case to the other according 
to Eq.~\eqref{eqn:ig_H}. It turns out that the phantom branch allows much larger values of the coupling $\xi$ and 
predicts a lower value for the Hubble constant without any 
prospect to reduce the $H_0$ tension. It is interesting to note that the extension of our study to the phantom case strengthen 
the correspondence between the kinetic term and the spatial curvature: the standard (phantom) kinetic term shifts the position 
to the right (left) as a negative (positive) spatial curvature.

\begin{figure}
\centering
\includegraphics[width=0.49\textwidth]{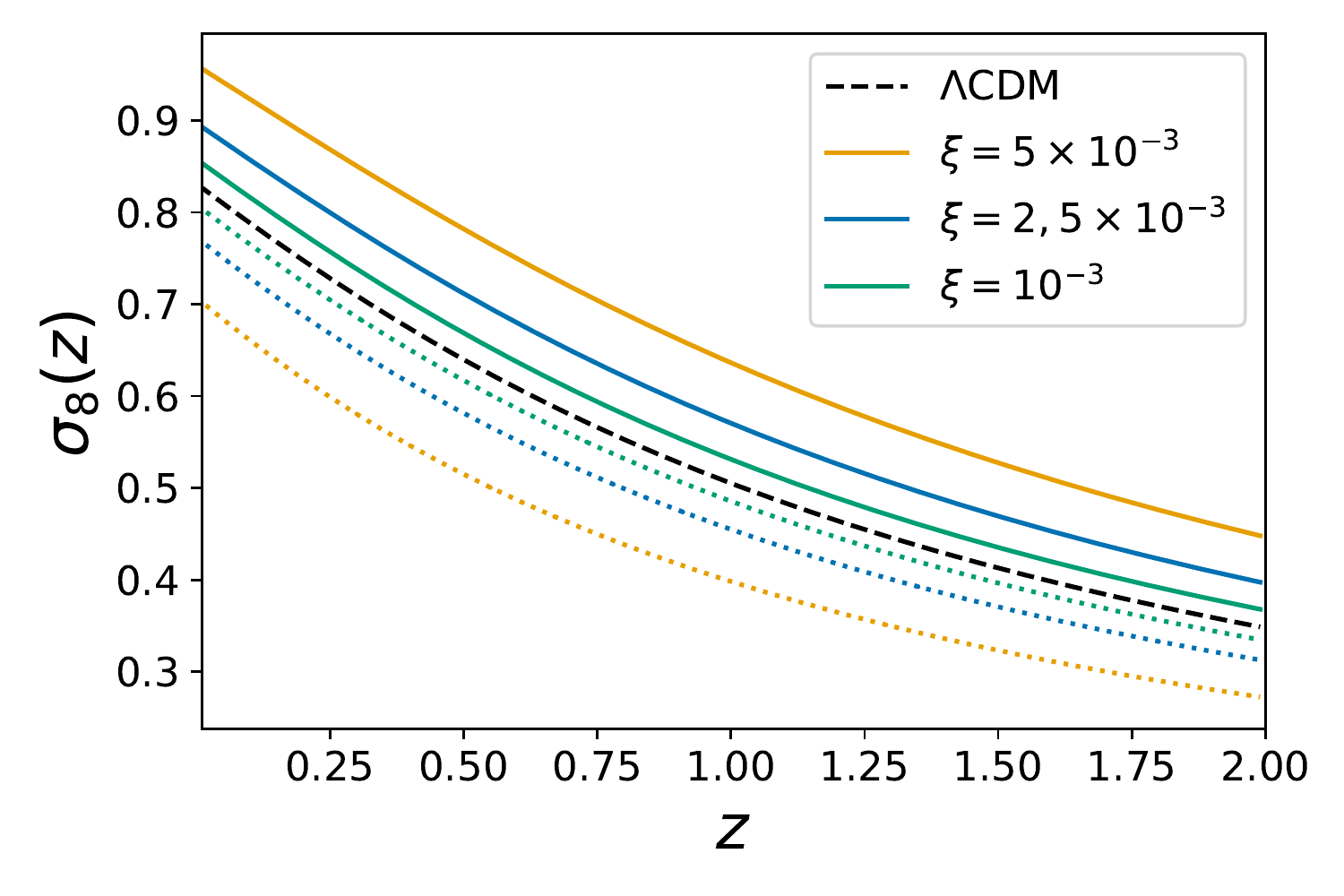}
\includegraphics[width=0.49\textwidth]{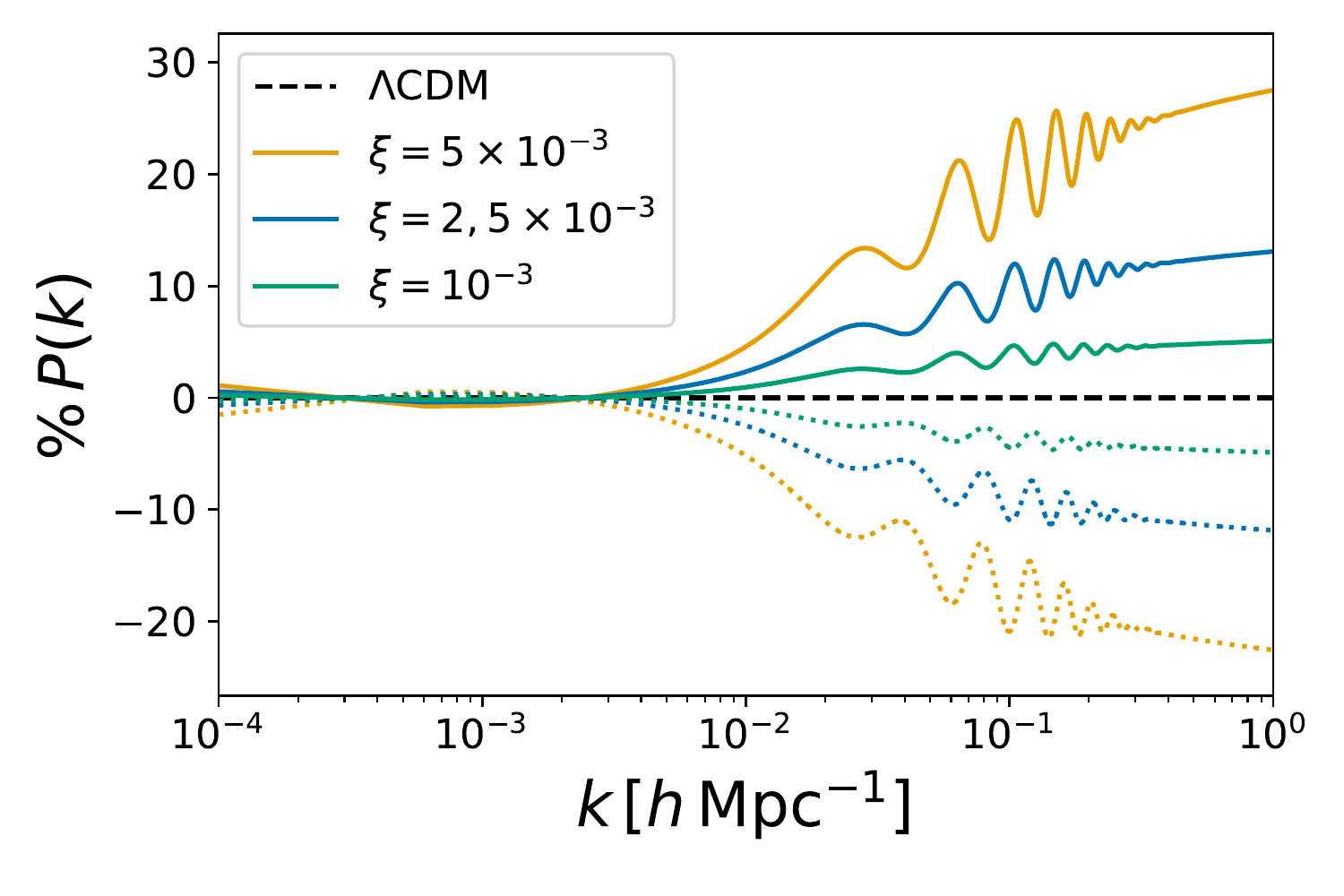}
\caption{Time evolution of the amplitude of matter perturbation within spheres of radius $8\,h^{-1}\,{\rm Mpc}$ (left panel) 
and relative differences of the linear matter power spectrum at $z=0$ with respect to $\Lambda$CDM (right panel) for 
different  values of the coupling parameter $\xi$ in the standard branch (solid lines) and in the phantom one (dashed lines) 
for IG ($F = \xi\sigma^2,\ V = \lambda F^2/4$).}
\label{fig:szandPk}
\end{figure}
Finally, it is interesting to note that the matter density root mean square fluctuations $\sigma_8$ goes toward lower 
values in the phantom branch compared to both the standard branch and the $\Lambda$CDM model predictions, see 
Fig.~\ref{fig:szandPk}. 
This behaviour can be understood studying the late-time solution of the perturbation equation for the matter density 
contrast in the linear regime, on sub-horizon scales
\begin{equation} \label{eqn:delta_m1}
    \delta_{\rm m}'' + \left(\frac{3}{a} + \frac{H'}{H}\right)\delta_{\rm m}' 
    - \frac{G_{\rm eff}}{2 G H^2} \frac{\rho_{\rm m}}{a^2} \delta_{\rm m} \simeq 0
\end{equation}
where primes are derivatives with respect to the scale factor $a$. 
By rewriting the Friedmann equations~\eqref{eqn:Friedmann1}-\eqref{eqn:Friedmann2}
\begin{align} 
    &H^2 = \frac{\rho + V}{3 F (1+f)} \\
    &\frac{H'}{H} = -\frac{3}{2a} - \frac{3}{a}w - \frac{F'}{2F} + \frac{f'}{2}
\end{align}
introducing the quantity 
\begin{equation}
    f = -a\frac{F_\sigma}{F}\sigma' - \frac{a^2 Z}{6F} \sigma'^2
\end{equation}
and where we used $\rho'/\rho = -3(1+w)/a$, we can write Eq.~\eqref{eqn:delta_m1} as
\begin{equation}
    \delta_{\rm m}'' + \left[\frac{3}{2a}(1 - w) - \frac{F'}{2F} + \frac{f'}{2}\right]\delta_{\rm m}' 
    - \frac{3}{2a^2} \frac{2ZF + 4 F_\sigma^2}{2ZF + 3 F_\sigma^2} (1 - f) \frac{\rho_{\rm m}}{\rho + V} 
    \delta_{\rm m} \simeq 0 \,.
\end{equation}
During the matter-dominated era, the scalar field evolves as $\sigma \sim a^{2Z\xi}$ \cite{Finelli:2007wb} leading for IG to
\begin{equation}
    f \sim -\frac{10Z\xi}{3}
\end{equation}
and consequently
\begin{equation} \label{eqn:delta_m2}
    \delta_{\rm m}'' + \frac{3}{2a}\left(1 - \frac{4Z\xi}{3}\right)\delta_{\rm m}' 
    - \frac{3}{2a^2} \left(1 + \frac{16Z\xi}{3}\right) \delta_{\rm m} \simeq 0 \,.
\end{equation}
In the weak coupling regime for $\xi \ll 1$, which turns out to be the range allowed from observations, the 
leading-order growing solution of Eq.~\eqref{eqn:delta_m2} goes as $\delta_{\rm m} \sim a^{1+4Z\xi}$ showing a slower 
(faster) growth of structures compared to the $\Lambda$CDM case for $Z < 0$ ($Z > 0$) during the matter dominated era. 

We show the results for most of the combination of datasets on Fig.~\ref{fig:ig_ph} (in Appendix~\ref{sec:app_bao}, we 
show a comparison between P18+BAO and P18+BAO+FS). The marginalized upper bound on the coupling parameter $\xi$ at 95\% 
CL corresponds to $< 0.0024$ for FS+SN, $< 0.0018$ for P18, $< 0.00046$ for P18+BAO, and $< 0.00040$ for P18+BAO+SN; 
see Tab.~\ref{tab:ig} for the constraints on all the parameters.
\begin{figure}
\centering
\includegraphics[width=0.98\textwidth]{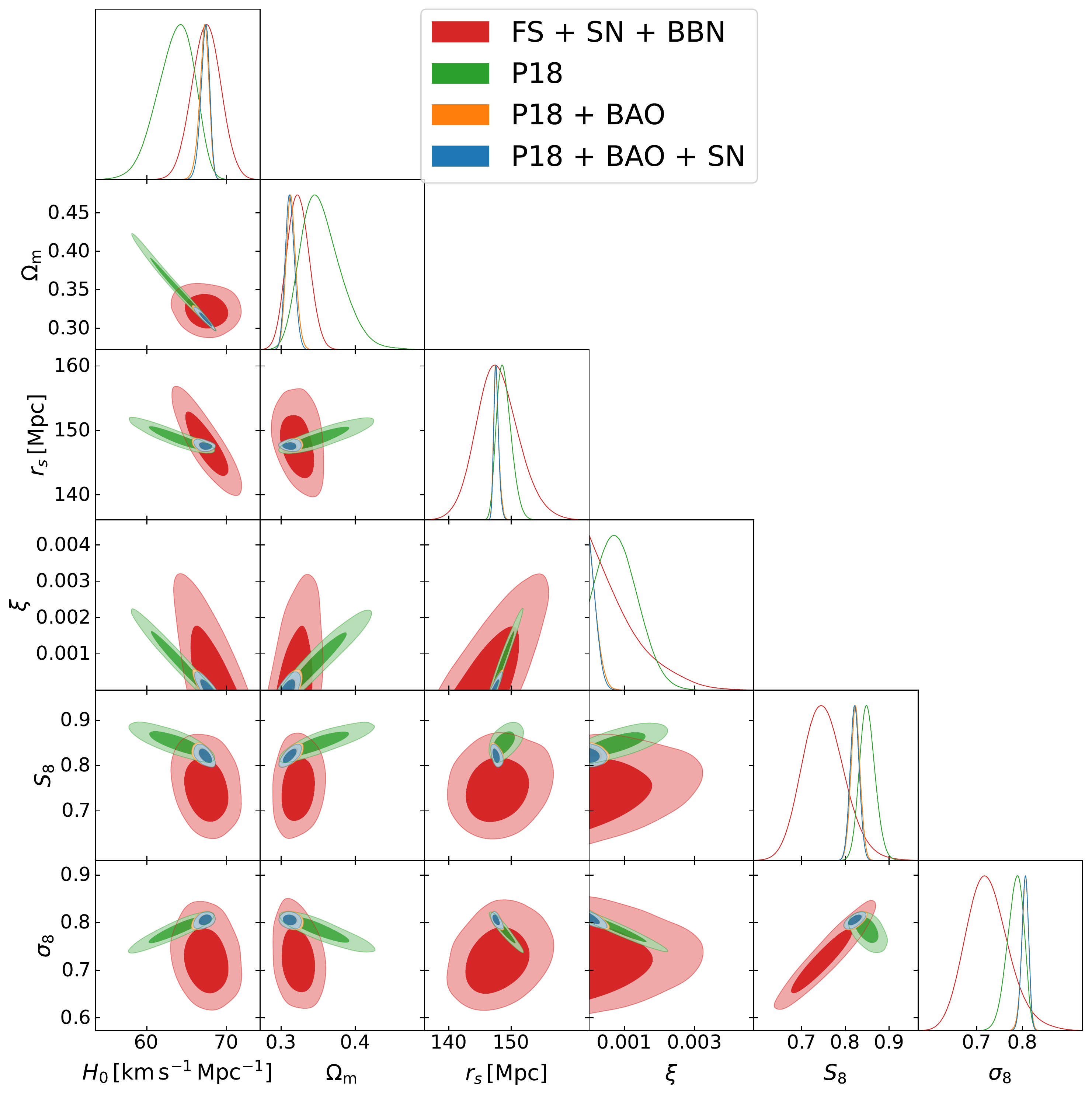}
\caption{Marginalized joint 68\% and 95\% CL regions 2D parameter space using the CMB-independent combination FS+SN (red), 
P18 (green), the combination P18+BAO (orange), and the combination P18+BAO+SN (blue) for IG ($F = \xi\sigma^2,\ V = \lambda F^2/4$) 
in the phantom branch ($Z = -1$).}
\label{fig:ig_ph}
\end{figure}

In Fig.~\ref{fig:ig_ph}, we see the larger marginalized uncertainties 
for the analysis without CMB information, i.e. combining FS with SN and a Gaussian prior on $\omega_{\rm b}$ motivated from 
BBN, and the analysis with CMB alone. In these cases larger value of $\xi$ can be accommodated by changes in the density 
parameters and in the scalar spectral index.

The marginalized means and uncertainties for the Hubble constant $H_0\,\left[{\rm km}\,{\rm s}^{-1}\,{\rm Mpc}^{-1}\right]$ 
at 68\% CL correspond to $67.4\pm 1.8$ for FS+SN, $63.6^{+2.7}_{-1.9}$ for P18, $67.17^{+0.64}_{-0.50}$ for P18+BAO, 
and $67.29^{+0.60}_{-0.47}$ for P18+BAO+SN; all of them are lower than the corresponding results we found in the standard 
branch, see Refs.~\cite{Ballardini:2020iws,Ballardini:2021eox}.
The upper bound on $\xi$ becomes much tighter, i.e. $\xi < 0.00016$ at 95\% CL, once we add a prior on $H_0$ in order to 
reproduce a larger value of the Hubble parameter, i.e. $68.34\pm 0.41$ at 68\% CL, see Fig.~\ref{fig:ig_ph_H0} and 
Tab.~\ref{tab:H0}.

The marginalized constraint on the present value of $\sigma_8$ at 68\% CL corresponds to 
$0.717\pm 0.049$ for FS+SN, $0.784^{+0.021}_{-0.015}$ for P18, $0.799^{+0.010}_{-0.009}$ for P18+FS, and $0.8059\pm 0.0058$ 
for P18+FS+SN. 
However, the combination $S_8 \equiv \sigma_8 \sqrt{\Omega_{\rm m}/0.3}$, commonly used to quantify the tension between 
{\em Planck} and weak lensing of galaxies measurements, moves to the wrong direction.
In order not to spoil the fit to CMB and galaxy measurements, an increase of the matter energy density is needed to 
compensate for the shifted position of the CMB acoustic peaks and of the BAO. 
Indeed, we find for $S_8$ $0.744\pm 0.050$ for FS+SN, $0.850^{+0.016}_{-0.019}$ for P18, 
$0.831^{+0.011}_{-0.012}$ for P18+FS, and $0.825\pm 0.012$ for P18+FS+SN at 68\% CL, resulting to be larger than in 
the standard branch as shown in Fig.~\ref{fig:ig_ph_comp}.

The constraints found are at odds with the parameter space free from ghost which corresponds for 
IG to $\xi > 1/6$ according to Eqs.~\eqref{eqn:ghostfree1}-\eqref{eqn:ghostfree2}. 
Note that, this condition for $\xi$ can be relaxed if one considers a more general Lagrangian with respect to the one introduced 
in Eq.~\eqref{eqn:action}. Higher order terms in the kinetic energy  $X \equiv - \partial_\mu \sigma \partial^\mu \sigma / 2$ \cite{Garriga:1999vw} appear in low-energy effective string theory \cite{Gasperini:2002bn} or in tachyon condensation \cite{Arkani-Hamed:2003pdi}. In a more general Lagrangian containing also a Galileon term $G_3$ \cite{Nicolis:2008in,Deffayet:2009wt} as in 
${\cal L} = G_4(\sigma)R + G_2(\sigma,X) + G_3(\sigma,X)\square\sigma$, the conditions for the avoidance of such instabilities are
\begin{equation} \label{eqn:noghostG3}
q_s\equiv \,4 G_4 \big\{ G_{2X} + 2 G_{3\sigma} +\dot\sigma \big[(G_{2XX} + G_{3X\sigma})\dot\sigma -6 G_{3X} H \big] \big\} + 3 (2G_{4\sigma} + G_{3X}\dot\sigma^2)^2 > 0
\end{equation}
\begin{align} \label{eqn:noLaplG3}
\begin{split}
c_s^2\equiv \,\big[&4 G_{2X} G_4 +8 G_{3\sigma} G_{4} + (6 G_{4\sigma}^2 - G_{3X}\dot\sigma^2) (2 G_{4\sigma}^2 + G_{3X}\dot\sigma^2) \\
&-8 G_4 (G_{3X}\ddot\sigma + 2G_{3X}H\dot\sigma + G_{3X\sigma}\dot\sigma^2) \big]/q_s > 0,
\end{split}
\end{align}
which reduce to Eqs.~\eqref{eqn:ghostfree1}-\eqref{eqn:ghostfree2} when $G_3=0$; but, in general, depending on the 
functional form of the cubic interaction term and its magnitude, can allow for $\xi < 1/6$ while maintaining the theory 
free of ghost and Laplacian instabilities \cite{Silva:2009km,Kobayashi:2010wa,Zumalacarregui:2020cjh, Ferrari:2023xyz}.

\subsection{Phantom non-minimal coupling}
For NMC+ (NMC-) \cite{Rossi:2019lgt}, with coupling $F(\sigma) = N_{\rm Pl}^2 + \xi\sigma^2$ and coupling 
$V(\sigma) = \lambda F^2(\sigma)/4$, we sample on the dimensionless parameter 
$\Delta \tilde{N}_{\rm Pl} \equiv N_{\rm Pl}/M_{\rm Pl} - 1$ 
and $\xi$.

\begin{figure}
\centering
\includegraphics[width=0.98\textwidth]{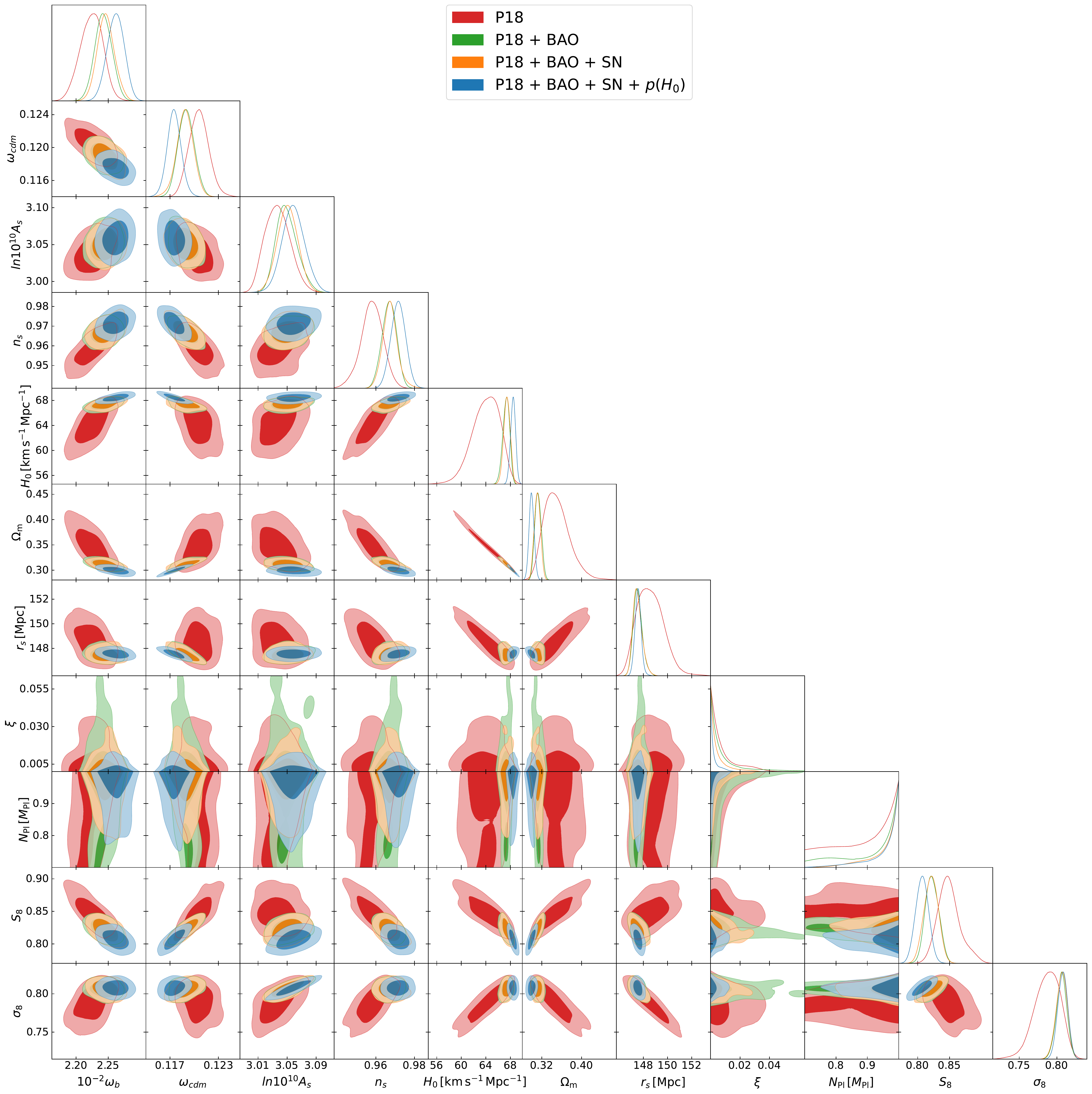}
\caption{Marginalized joint 68\% and 95\% CL regions 2D parameter space using using the P18 (green), the combination 
P18+BAO (orange), and the combination P18+BAO+SN (blue) for NMC+ ($F = N_{\rm Pl}^2 + \xi\sigma^2,\ V = \lambda F^2/4$) in 
the phantom branch ($Z = -1$).}
\label{fig:nmc_ph_xp}
\end{figure}

\begin{figure}
\centering
\includegraphics[width=0.98\textwidth]{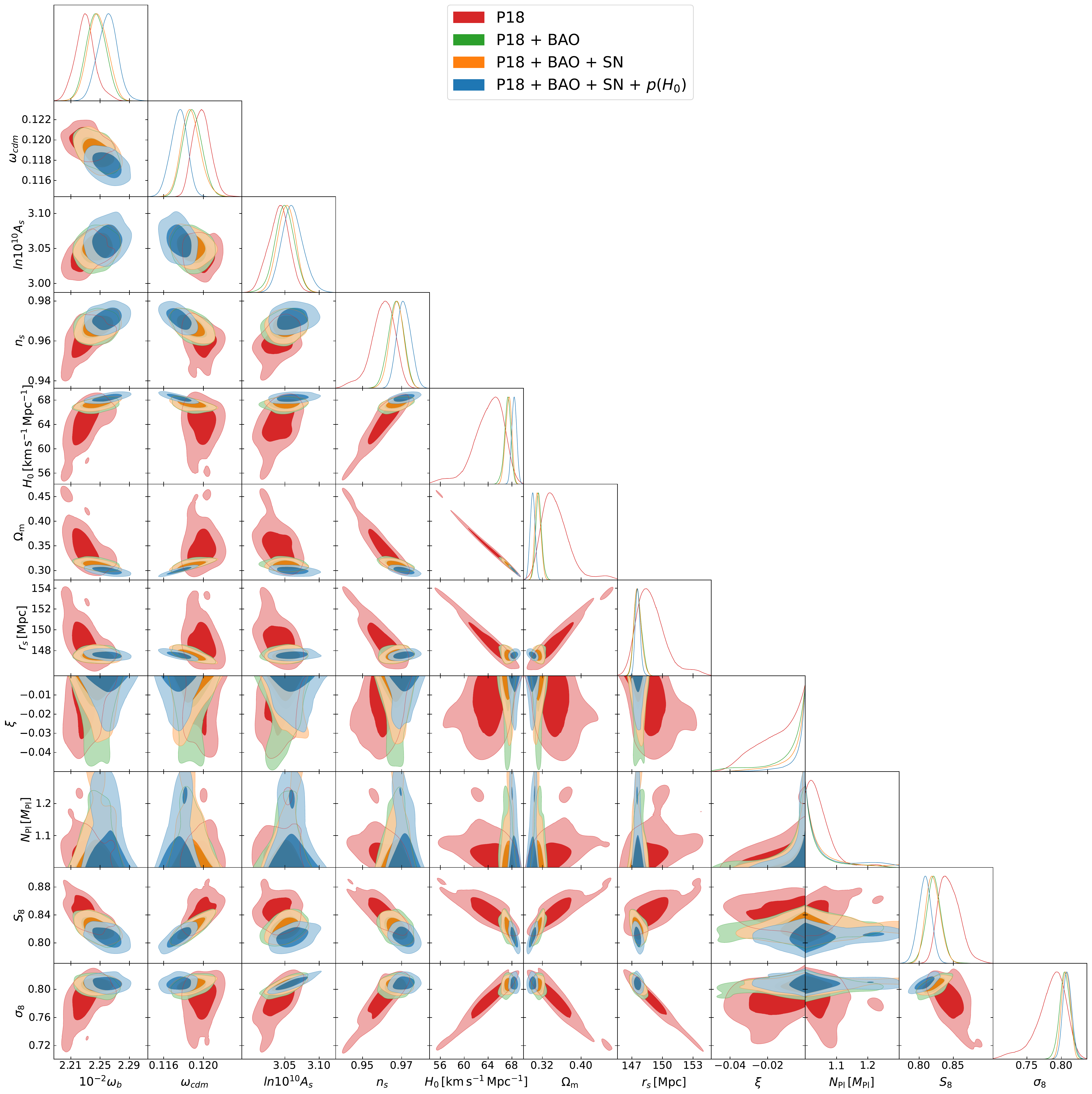}
\caption{Marginalized joint 68\% and 95\% CL regions 2D parameter space using using the P18 (red), the combination 
P18+BAO (green), the combination P18+BAO+SN (orange), and the combination P18+BAO+SN+$p(H_0)$ (blue) for NMC- 
($F = N_{\rm Pl}^2 + \xi\sigma^2,\ V = \lambda F^2/4$) in the phantom branch ($Z = -1$).}
\label{fig:nmc_ph_xm}
\end{figure}

We show the results for all the combinations of datasets on Fig.~\ref{fig:nmc_ph_xp} for NMC+ and for NMC- in 
Fig.~\ref{fig:nmc_ph_xm}. Marginalized constraints on cosmological parameters are consistent to the results obtained for 
IG for each combination of datasets.
The marginalized limits on the coupling parameters for NMC+ (NMC-) at 95\% CL correspond to $\xi < 0.0015$ ($>-0.039$) 
and $N_{\rm Pl} > 0.91$ ($< 1.18$) for P18+BAO, and to $\xi < 0.0019$ ($>-0.027$) and $N_{\rm Pl} > 0.83$ ($< 1.21$) 
for P18+BAO+SN; see Tabs.~\ref{tab:nmc_xp} and \ref{tab:nmc_xm} for the constraints on all the parameters.
As observed in Refs.~\cite{Rossi:2019lgt,Ballardini:2020iws}, there is a strong degeneracy between the coupling 
parameters $N_{\rm Pl}$ and $\xi$ for the form of non-minimal coupling $F(\sigma) = N_{\rm Pl}^2 + \xi\sigma^2$. Since 
cosmological observables are affected by contributions ${\cal O}\left(\xi\sigma^2/ N_{\rm Pl}^2\right)$, it is possible 
to compensate the effects due to a large value of $|\xi|$ increasing $|\tilde{N}_{\rm Pl}-1|$ and vice versa.

In this case Eqs.~\eqref{eqn:ghostfree1}-\eqref{eqn:ghostfree2} reduce to
\begin{equation} \label{eqn:noghost_nmc}
    \left(\frac{N_{\rm Pl}}{\xi\sigma}\right)^2 + \frac{1}{\xi} < 6 \,.
\end{equation}
Also in this case, a large portion of the allowed parameter space is at odds with Eq.~\eqref{eqn:noghost_nmc} despite the 
larger number of degrees of freedom.

\subsection{Phantom early modified gravity}
For EMG \cite{Braglia:2020auw}, with coupling $F(\sigma) = M_{\rm Pl}^2 + \xi\sigma^2$ and potential 
$V(\sigma) = \Lambda + \lambda\sigma^4/4$, we sample on the quantity $\xi$ 
and $V_0$ 
where $\lambda \equiv -10^{2 V_0}/M_{\rm Pl}^4$. In this case the scalar field decays into the 
minimum of the potential, i.e. $\sigma = 0$, and we do not have to impose the boundary condition \eqref{eqn:boundary} 
being automatically satisfied for each initial value of scalar field. 
This leads to a third free parameter which we identify with the initial value of the scalar field $\sigma_{\rm ini}$ 
$\left[M_{\rm Pl}\right]$.

\begin{figure}
\centering
\includegraphics[width=0.48\textwidth]{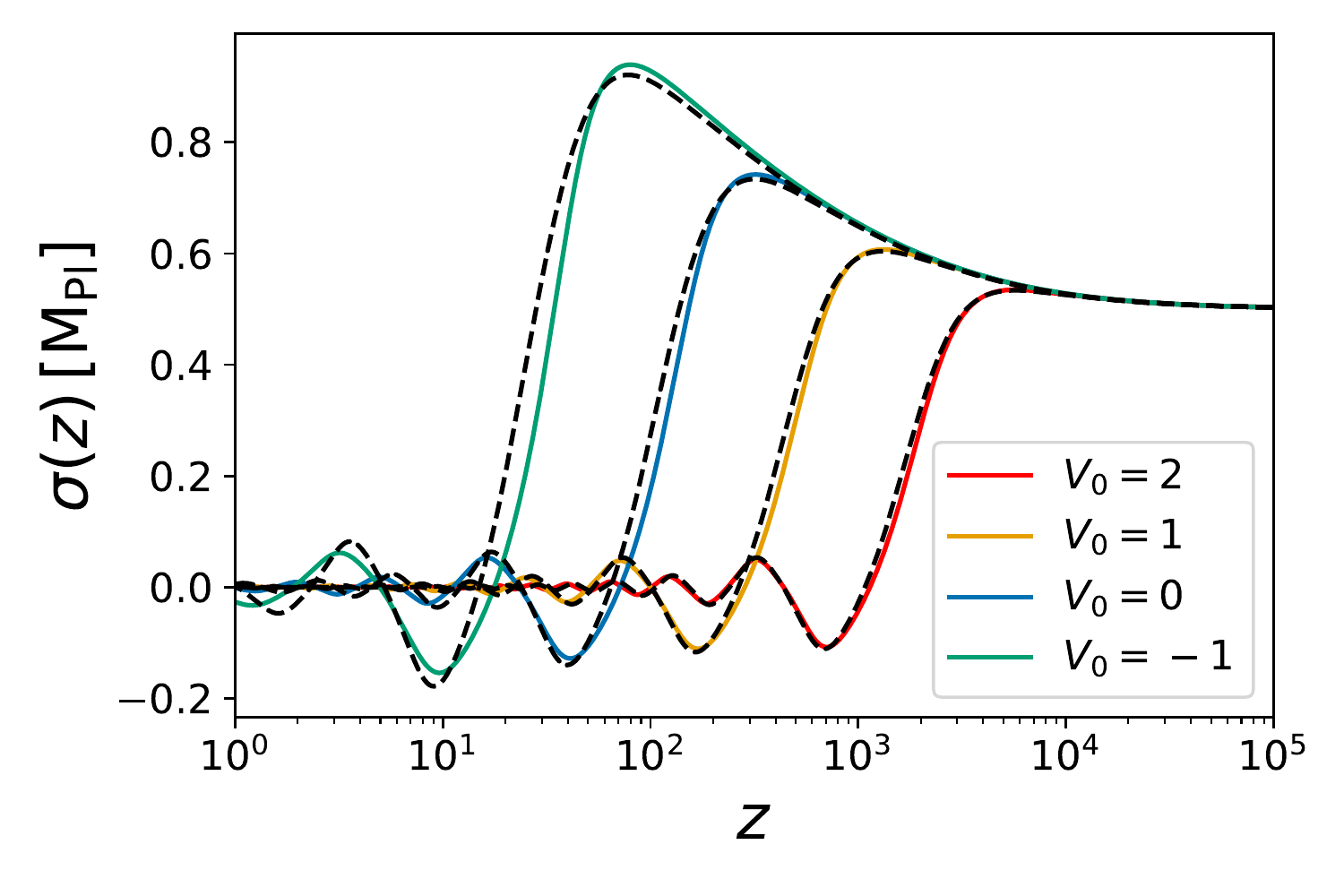}
\includegraphics[width=0.48\textwidth]{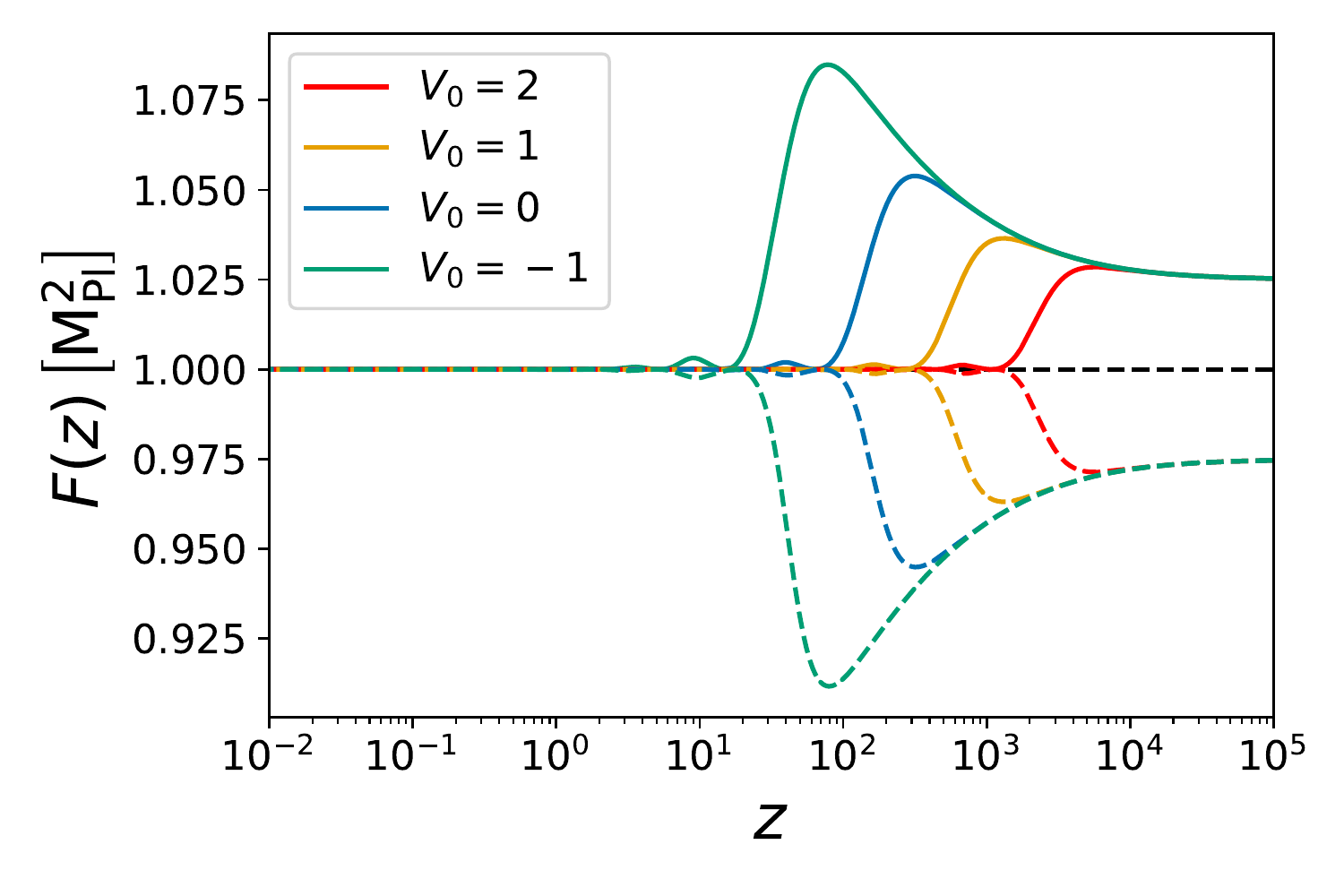}
\includegraphics[width=0.48\textwidth]{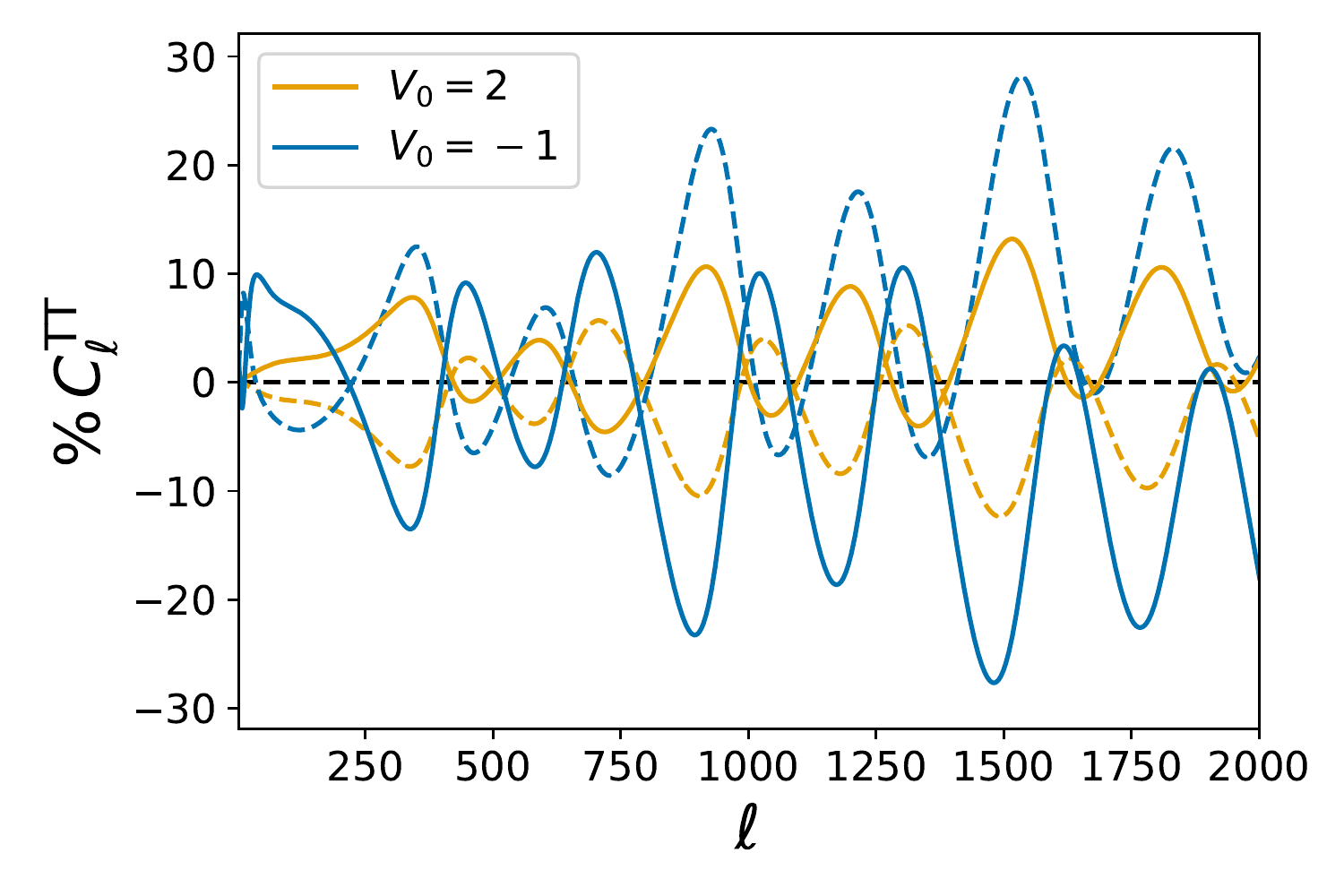}
\includegraphics[width=0.48\textwidth]{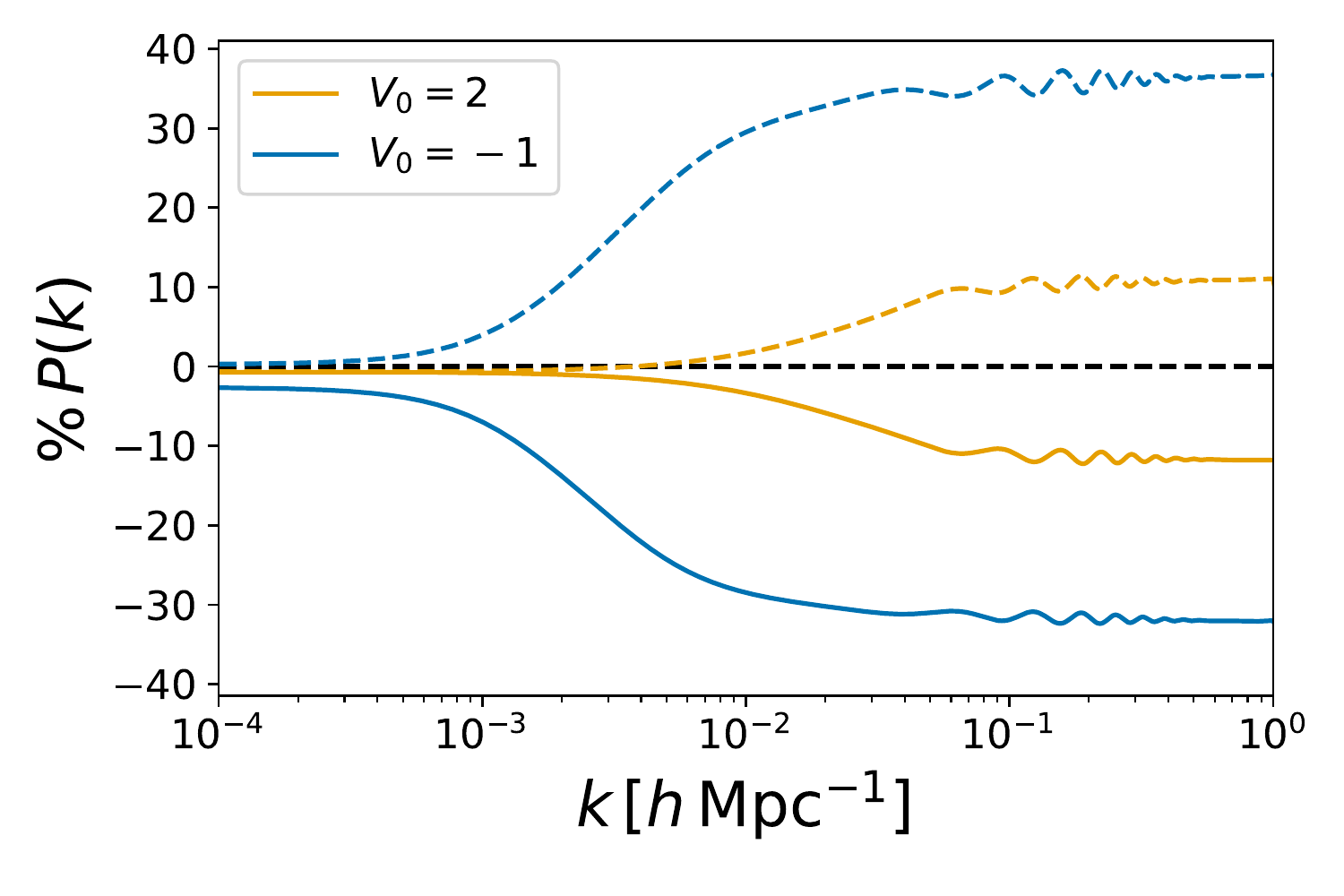}
\caption{Time evolution of the scalar field $\sigma$ (upper left panel) and of the coupling to the Ricci scalar $F(\sigma)$ 
(upper right panel). Relative differences of the CMB temperature anisotropies power spectrum with respect to the 
$\Lambda$CDM case (bottom left panel) and of the linear matter power spectrum at $z = 0$ (bottom right panel). 
Different lines correspond to different value of the amplitude of the effective potential $V_0$ for $|\xi| = 0.1$ in the 
standard branch (solid lines) and in the phantom one (dashed lines) for EMG 
($F = M_{\rm Pl}^2 + \xi\sigma^2,\, V = \Lambda + \lambda\sigma^4/4$).}
\label{fig:emg_bkg}
\end{figure}

In Fig.~\ref{fig:emg_bkg}, we compare the background evolution and spectra of EMG in the standard branch ($Z = 1$) to 
the phantom branch case ($Z = -1$). The evolution of the scalar field $\sigma$ is very similar in the two cases. Starting 
with the scalar field at rest in the radiation era, it starts to grow around the recombination driven by the coupling to 
the non-relativistic matter component and it is subsequently driven into damped coherent oscillations around the minimum 
of the quartic potential.
The different evolution of the coupling function, which increase in the branch with standard kinetic term while decreasing 
in the phantom branch before the scalar field starts to decay, due to the different sign of the coupling parameter $\xi$ 
induces different effects on the spectra. 
The acoustic peaks of the CMB temperature anisotropies angular power spectrum are shifted to right in the standard branch 
($Z = 1$) when the scalar field starts to move before recombination ($V_0 = 2$) and in the phantom branch ($Z = -1$) if 
the scalar field decays after recombination ($V_0 = -1$), vice versa they shift to the left with respect the $\Lambda$CDM case.
The situation is different on the linear matter power spectrum were we observe a suppression of power in the standard 
branch ($Z = 1$) despite the value of $V_0$ and an increase of power at small scales in the phantom branch ($Z = -1$). 
This highlights the importance of the combination of combining early- and late-time probes in order to break the degeneracy 
between the extra parameters of the model and also to discriminate between the two different branches.

\begin{figure}
\centering
\includegraphics[width=0.98\textwidth]{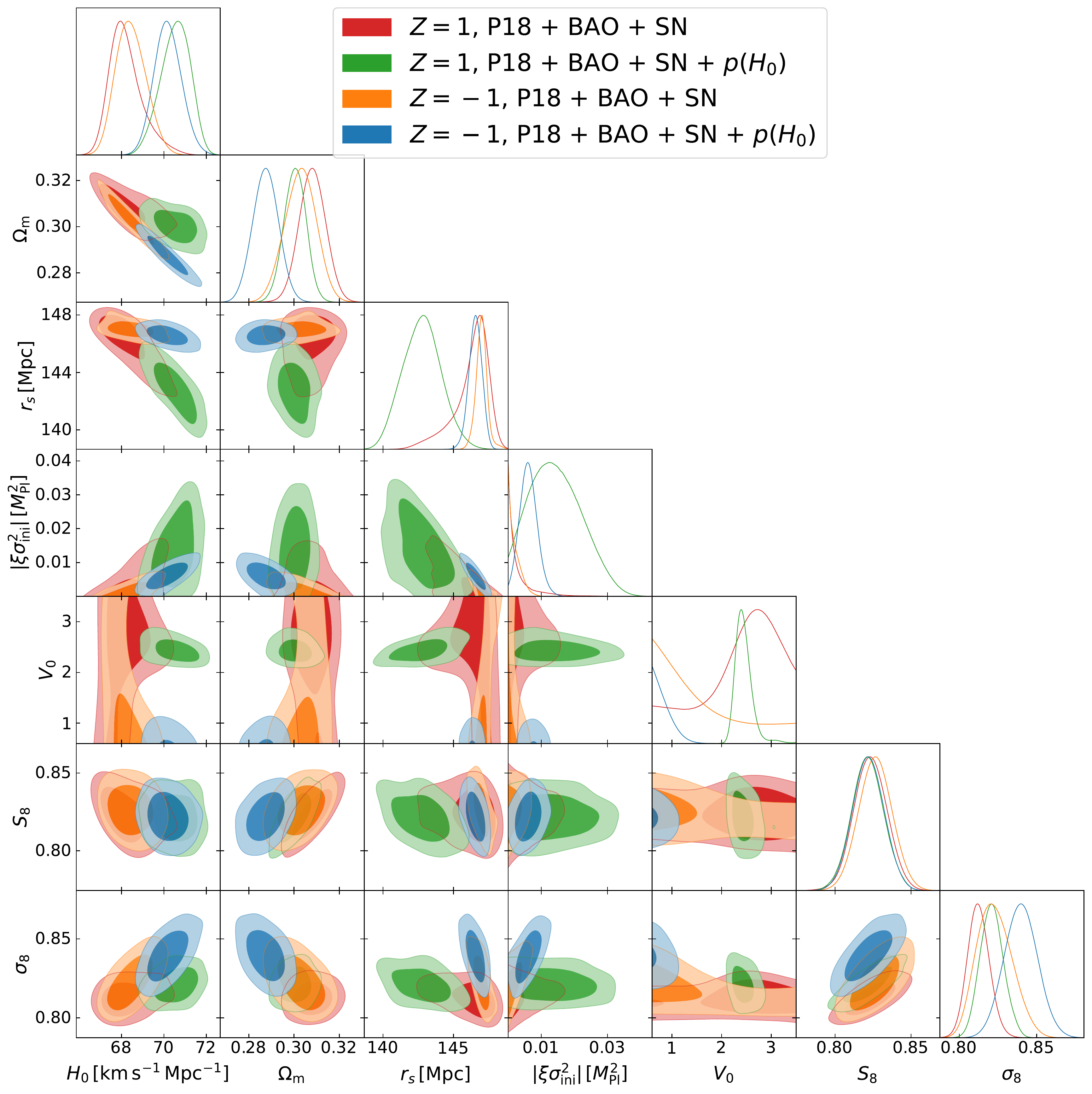}
\caption{Marginalized joint 68\% and 95\% CL regions 2D parameter space using P18+BAO+SN (orange) and the combination 
P18+BAO+SN+$p(H_0)$ (blue) for EMG ($F = M_{\rm Pl}^2 + \xi\sigma^2,\ V = \Lambda + \lambda\sigma^4/4$) in the phantom 
branch ($Z = -1$).}
\label{fig:emg_ph}
\end{figure}

We show the results for the combinations of datasets P18+BAO+SN and P18 +BAO+SN+$p(H_0)$ on Fig.~\ref{fig:emg_ph}. 
In this case we find that $\xi$ is not constrained by data on the prior considered $[-0.1,\,0]$. For this reason,  we show 
constraints on the combination $\xi\sigma_{\rm ini}^2$ $[M_{\rm Pl}^2]$ (connected to the additional contribution to the 
expansion rate evolution \eqref{eqn:ig_H} before recombination). 
The marginalized upper bound on the coupling combination $\xi\sigma_{\rm ini}^2$ at 95\% CL corresponds to $> -0.0026$ for 
P18+BAO+SN and when we include the Gaussian prior on the Hubble parameter we obtain at 95\% CL $-0.006 \pm 0.005$. 
Analogously, for the initial value of the scalar field $\sigma_{\rm ini}$ $[M_{\rm Pl}]$ we find $< 0.45$ for P18+BAO+SN 
and $0.35^{+0.17}_{-0.15}$ for P18+BAO+SN+$p(H_0)$ both at 95\% CL.
Also $V_0$ is not well constrained. We find a 95\% CL upper bound only when we include the Gaussian prior on the Hubble 
parameter corresponding to $V_0 < 0.81$.

The marginalized means and uncertainties for the Hubble constant $H_0\,\left[{\rm km}\,{\rm s}^{-1}\,{\rm Mpc}^{-1}\right]$ 
at 68\% CL correspond to $68.44^{+0.62}_{-0.79}$ for P18+BAO+SN and $70.18^{+0.59}_{-0.68}$ for P18+BAO+SN +$p(H_0)$. 
The marginalized constraints on $S_8$ correspond to $S_8 = 0.827\pm 0.011$ for P18+BAO+SN and $S_8 = 0.822\pm 0.010$ for 
P18+BAO+SN+$p(H_0)$. See Tab.~\ref{tab:emg} for the constraints on all the parameters.

\section{Conclusions} \label{sec:conclusions}
We have studied the dynamics and inferred the cosmological constraints for modified gravity models with a nonminimally 
coupled scalar with a kinetic term which can also have a negative sign.
For stable models with an effectively massless scalar field $\sigma$, like IG and NMC, the change of sign in front of the kinetic term 
of the scalar field modifies the evolution of the scalar field which is at rest during the radiation-dominated 
epoch and evolves like $\sigma \sim a^{2Z\xi}$ during the matter-dominated era.

We have shown the effect of the sign of the kinetic term on cosmological observables.
We have computed the marginalized constraints for different combination of cosmological datasets by allowing the coupling to 
the Ricci scalar and the rest of cosmology (standard cosmological parameters and nuisance ones) to vary. 
Combining {\em Planck} 2018 DR3 measurements with BAO from BOSS and eBOSS, and uncalibrated SN I$\alpha$ from the Pantheon 
sample we constrain the coupling parameters at 95\% CL to $\xi < 0.00040$ for $F(\sigma)=\xi\sigma^2$ and for 
$F(\sigma)=N_{\rm Pl}^2+\xi\sigma^2$ to $\xi < 0.0019$ ($>-0.027$) and $N_{\rm Pl} > 0.83$ ($< 1.21$).

Nonminimally coupled scalar-tensor theories with early-time deviation from GR predictions usually lead to higher 
values of the Hubble parameter $H_0$, a lower value of the matter density parameter $\Omega_{\rm m}$, and a larger 
value of the $\sigma_8$ \cite{SolaPeracaula:2020vpg,Zumalacarregui:2020cjh,Ballardini:2020iws}. In their phantom construction, the modified 
evolution of the scalar field, connected to a different time evolution of the effective gravitational constant, inverts 
the degeneracy between these parameters and the coupling ones. Indeed, we find a lower values of both $\sigma_8$ and 
$H_0$ compared to the branch with standard kinetic term.

We have also studied the phantom version of the EMG model introduced in Ref.~\cite{Braglia:2020auw}.
While the evolution of the scalar field is very similar, with the quartic potential leading the scalar field to 
decay into damped coherent oscillations, different signatures appear on the cosmological observables.
Compared to the $\Lambda$CDM model, the CMB acoustic peaks of the CMB are shifted to right in the standard branch 
($Z = 1$) when the scalar field starts to move before recombination ($V_0 = 2$) and in the phantom branch ($Z = -1$) 
if the scalar field decays after recombination ($V_0 = -1$), vice versa they shift to the left. 
Matter perturbations on sub-horizon scales are suppressed in the standard branch ($Z = 1$) and enhanced in the 
phantom branch ($Z = -1$) despite the value of amplitude of the self-interaction term parameterized by $V_0$.

The allowed parameter space for the coupling parameters by our analysis is at odds with the parameter space free 
from ghosts and Laplacian instabilities. It would be interesting to understand if instead there are healthy 
scalar-tensor theories which retain the possibility to alleviate the current tensions between different cosmological observations. 

\section*{Acknowledgments}
MB and FF acknowledge financial support from the INFN InDark initiative and from the COSMOS 
network (www.cosmosnet.it) through the ASI (Italian Space Agency) Grants 2016-24-H.0 
and 2016-24-H.1-2018, as well as 2020-9-HH.0 (participation in LiteBIRD phase A).
This work has made use of computational resources of INAF OAS Bologna and of the
CNAF HPC cluster in Bologna.

\newpage

\appendix
\section{Tables} \label{sec:app_tab}
\begin{table*}[h!]
{\small
\centering
\begin{tabular}{l|ccc}
\hline
\hline
                                         & P18 & P18 + BAO & P18 + BAO + SN  \\
\hline
$\omega_{\rm b}$                         & $0.02223\pm 0.00017$        & $0.02244\pm 0.00013$     & $0.02245\pm 0.00013$  \\
$\omega_{\rm c}$                         & $0.1204\pm 0.0012$          & $0.11896\pm 0.00099$       & $0.11882\pm 0.00096$  \\
$H_0$ [km s$^{-1}$Mpc$^{-1}$]            & $63.6^{+2.7}_{-1.9}$        & $67.17^{+0.65}_{-0.48}$  & $67.29^{+0.60}_{-0.47}$ \\
$\tau$                                   & $0.0523\pm 0.0071$          & $0.0584^{+0.0070}_{-0.0083}$       & $0.0584^{+0.0068}_{-0.0076}$  \\
$\ln \left(  10^{10} A_{\rm s} \right)$  & $3.037\pm 0.015$            & $3.051^{+0.014}_{-0.016}$         & $3.051\pm 0.014$  \\
$n_{\rm s}$                              & $0.9574^{+0.0067}_{-0.0057}$          & $0.9668\pm 0.0038$       & $0.9671\pm 0.0036$  \\
$\xi$                                                     & $< 0.0018$ (95\% CL)       & $< 0.00046$ (95\% CL)     & $< 0.00040$ (95\% CL)  \\
\hline
$\gamma_{\rm PN}$                                             & $> 0.9928$ (95\% CL)       & $> 0.9982$ (95\% CL)      & $> 0.9984$ (95\% CL) \\
$\delta G_\mathrm{N}/G_\mathrm{N}$ (z=0)                  & $< 0.057$ (95\% CL)       & $< 0.014$ (95\% CL) & $< 0.012$ (95\% CL)  \\
$10^{13} \dot{G}_\mathrm{N}/G_{\rm N}$ (z=0) [yr$^{-1}$]  & $< 2.43$ (95\% CL)        & $< 0.57$ (95\% CL)       & $< 0.50$ (95\% CL)  \\
$G_\mathrm{N}/G$ (z=0)                                    & $0.9982^{+0.0013}_{-0.00074}$  & $0.99965^{+0.00033}_{-0.00013}$          & $0.99968^{+0.00030}_{-0.00012}$  \\
\hline
$\Omega_{\rm m}$                        & $0.354^{+0.020}_{-0.032}$  & $0.3135^{+0.0059}_{-0.0068}$         & $0.3120^{+0.0056}_{-0.0065}$  \\
$\sigma_8$                              & $0.784^{+0.021}_{-0.015}$  & $0.8053^{+0.0083}_{-0.0069}$  & $0.8053^{+0.0078}_{-0.0066}$  \\
$S_8$                                   & $0.850^{+0.016}_{-0.019}$           & $0.823\pm 0.011$           & $0.821\pm 0.010$  \\
$r_s$ [Mpc]                             & $148.89^{+0.8}_{-1.4}$   & $147.62^{+0.31}_{-0.50}$      & $147.62^{+0.29}_{-0.44}$  \\
\hline
$\Delta \chi^2$                         & $-2.8$ & $0$  & $-0.5$  \\
\hline
\end{tabular}}
\caption{\label{tab:ig} 
Constraints on the main and derived parameters (at 68\% CL if not otherwise stated) considering 
P18 in combination with BAO and BAO+SN for the {\bf IG} model.}
\end{table*}

\begin{table*}[h!]
{\small
\centering
\begin{tabular}{l|ccc}
\hline
\hline
                                         & P18 & P18 + BAO & P18 + BAO + SN  \\
\hline
$\omega_{\rm b}$                         & $0.02224^{+0.00018}_{-0.00016}$        & $0.02246\pm0.00013$     & $0.02246^{+0.00011}_{-0.00014}$  \\
$\omega_{\rm c}$                         & $0.1206\pm 0.0012$          & $0.1190^{+0.0014}_{-0.0011}$       & $0.1189\pm 0.0010$  \\
$H_0$ [km s$^{-1}$Mpc$^{-1}$]            & $64.1^{+2.6}_{-1.7}$        & $67.28\pm 0.59$  & $67.42\pm 0.52$ \\
$\tau$                                   & $0.0514\pm 0.0081$          & $0.0590\pm 0.0052$       & $0.0583\pm 0.0071$  \\
$\ln \left(  10^{10} A_{\rm s} \right)$  & $3.037^{+0.015}_{-0.023}$            & $3.0517^{+0.0074}_{-0.015}$         & $3.051\pm 0.014$  \\
$n_{\rm s}$                              & $0.9580^{+0.0058}_{-0.0047}$          & $0.9673\pm 0.0042$       & $0.9674\pm 0.0039$  \\
$\xi$                                                     & $< 0.030$ (95\% CL)       & $< 0.015$ (95\% CL)     & $< 0.019$ (95\% CL)  \\
$N_{\rm Pl}$ [$M_{\rm Pl}$]    &   $-$  &   $> 0.91$ (95\% CL)  &   $> 0.83$ (95\% CL)  \\
\hline
$\gamma_{\rm PN}$                                             & $> 0.9941$ (95\% CL)       & $> 0.9986$ (95\% CL)      & $> 0.9987$ (95\% CL) \\
$\beta_{\rm PN}$                                             & $> 0.999965$ (95\% CL)       & $> 0.999994$ (95\% CL)      & $> 0.999994$ (95\% CL) \\
$\delta G_\mathrm{N}/G_\mathrm{N}$ (z=0)                  & $< 0.052$ (95\% CL)       & $< 0.011$ (95\% CL) & $< 0.012$ (95\% CL)  \\
$10^{13} \dot{G}_\mathrm{N}/G_{\rm N}$ (z=0) [yr$^{-1}$]  & $< 1.94$ (95\% CL)        & $< 0.42$ (95\% CL)       & $< 0.40$ (95\% CL)  \\
$G_\mathrm{N}/G$ (z=0)                                    & $1.00134^{+0.00063}_{-0.0011}$  & $1.000255^{+0.000097}_{-0.00024}$          & $1.000221^{+0.000084}_{-0.00023}$  \\
\hline
$\Omega_{\rm m}$                        & $0.349^{+0.017}_{-0.030}$  & $0.3121^{+0.0068}_{-0.0056}$         & $0.3110\pm 0.0061$  \\
$\sigma_8$                              & $0.788^{+0.021}_{-0.013}$  & $0.8069\pm 0.0069$  & $0.8064^{+0.0079}_{-0.0065}$  \\
$S_8$                                   & $0.849^{+0.013}_{-0.019}$           & $0.823^{+0.014}_{-0.009}$           & $0.821^{+0.012}_{-0.011}$  \\
$r_s$ [Mpc]                             & $148.56^{+0.90}_{-1.3}$   & $147.54^{+0.30}_{-0.48}$      & $147.52^{+0.27}_{-0.44}$  \\
\hline
$\Delta \chi^2$                         & $-1.5$ & $0$  & $-0.5$  \\
\hline
\end{tabular}}
\caption{\label{tab:nmc_xp} 
Constraints on the main and derived parameters (at 68\% CL if not otherwise stated) considering 
P18 in combination with BAO and BAO+SN for the {\bf NMC+} model.}
\end{table*}

\begin{table*}[h!]
{\small
\centering
\begin{tabular}{l|ccc}
\hline
\hline
                                         & P18 & P18 + BAO & P18 + BAO + SN  \\
\hline
$\omega_{\rm b}$                         & $0.02230\pm 0.00014$        & $0.02245\pm 0.00013$     & $0.02247\pm 0.00013$  \\
$\omega_{\rm c}$                         & $0.11982^{+0.00068}_{-0.0011}$          & $0.11891\pm 0.00094$       & $0.11875^{+0.00078}_{-0.0010}$  \\
$H_0$ [km s$^{-1}$Mpc$^{-1}$]            & $64.1^{+3.1}_{-2.1}$        & $67.26^{+0.59}_{-0.45}$  & $67.44^{+0.57}_{-0.45}$ \\
$\tau$                                   & $0.0548^{+0.0072}_{-0.0059}$          & $0.0573^{+0.0061}_{-0.0074}$       & $0.0590\pm 0.0068$  \\
$\ln \left(  10^{10} A_{\rm s} \right)$  & $3.041^{+0.017}_{-0.013}$            & $3.049\pm 0.014$         & $3.052^{+0.014}_{-0.012}$  \\
$n_{\rm s}$                              & $0.9604^{+0.0067}_{-0.0045}$          & $0.9669^{+0.0043}_{-0.0035}$       & $0.9675\pm 0.0036$  \\
$\xi$                                    & $> -0.036$ (95\% CL)       & $> -0.039$ (95\% CL)     & $> -0.027$ (95\% CL)  \\
$N_{\rm Pl}$ [$M_{\rm Pl}$]              & $< 1.13$ (95\% CL)   &   $< 1.18$ (95\% CL)  &   $< 1.21$ (95\% CL)  \\
\hline
$\gamma_{\rm PN}$                                         & $> 0.988$ (95\% CL)       & $> 0.998$ (95\% CL)      & $> 0.998$ (95\% CL) \\
$\beta_{\rm PN}$                                          & $< 1.00018$ (95\% CL)       & $< 1.000022$ (95\% CL)      & $< 1.000017$ (95\% CL) \\
$\delta G_\mathrm{N}/G_\mathrm{N}$ (z=0)                  & $< 0.060$ (95\% CL)       & $< 0.012$ (95\% CL) & $< 0.010$ (95\% CL)  \\
$10^{13} \dot{G}_\mathrm{N}/G_{\rm N}$ (z=0) [yr$^{-1}$]  & $< 3.85$ (95\% CL)        & $< 0.62$ (95\% CL)       & $< 0.50$ (95\% CL)  \\
$G_\mathrm{N}/G$ (z=0)                                    & $1.00224^{+0.00080}_{-0.0021}$  & $1.00037^{+0.00012}_{-0.00037}$          & $1.00030^{+0.00012}_{-0.00030}$  \\
\hline
$\Omega_{\rm m}$                        & $0.348^{+0.021}_{-0.033}$  & $0.3125^{+0.0052}_{-0.0065}$         & $0.3106^{+0.0050}_{-0.0068}$  \\
$\sigma_8$                              & $0.786^{+0.025}_{-0.011}$  & $0.8047\pm 0.0076$  & $0.8064\pm 0.0062$  \\
$S_8$                                   & $0.844^{+0.011}_{-0.018}$           & $0.821\pm 0.011$           & $0.8204^{+0.0091}_{-0.012}$  \\
$r_s$ [Mpc]                             & $148.91^{+0.77}_{-1.6}$   & $147.60^{+0.28}_{-0.43}$      & $147.57^{+0.30}_{-0.40}$  \\
\hline
$\Delta \chi^2$                         & $-2.8$ & $0$  & $-0.3$  \\
\hline
\end{tabular}}
\caption{\label{tab:nmc_xm} 
Constraints on the main and derived parameters (at 68\% CL if not otherwise stated) considering 
P18 in combination with BAO and BAO+SN for the {\bf NMC-} model.}
\end{table*}

\begin{table*}[h!]
{\small
\centering
\begin{tabular}{l|ccc}
\hline
\hline
                                         & IG & NMC+ & NMC-  \\
\hline
$\omega_{\rm b}$                         & $0.02260^{+0.00012}_{-0.00014}$        & $0.02262\pm 0.00013$     & $0.02260^{+0.00014}_{-0.00012}$  \\
$\omega_{\rm c}$                         & $0.11747\pm 0.00086$          & $0.11752\pm 0.00086$       & $0.11753^{+0.00095}_{-0.00069}$  \\
$H_0$ [km s$^{-1}$Mpc$^{-1}$]            & $68.34\pm 0.41$        & $68.42^{+0.44}_{-0.36}$  & $68.38\pm 0.41$ \\
$\tau$                                   & $0.0617^{+0.0067}_{-0.0085}$          & $0.0636^{+0.0071}_{-0.0081}$       & $0.0644^{+0.0068}_{-0.0090}$  \\
$\ln \left(  10^{10} A_{\rm s} \right)$  & $3.055^{+0.013}_{-0.017}$            & $3.059^{+0.014}_{-0.016}$         & $3.061^{+0.014}_{-0.017}$  \\
$n_{\rm s}$                              & $0.9711\pm 0.0036$          & $0.9716\pm 0.0037$       & $0.9712\pm 0.0035$  \\
$\xi$                                    & $< 0.000075$ (95\% CL)       & $< 0.0096$ (95\% CL)     & $> -0.022$ (95\% CL)  \\
$N_{\rm Pl}$ [$M_{\rm Pl}$]              & $0$   &   $> 0.82$ (95\% CL)  &   $< 1.24$ (95\% CL)  \\
\hline
$\gamma_{\rm PN}$                                         & $> 0.9993$ (95\% CL)       & $> 0.9995$ (95\% CL)      & $> 0.9994$ (95\% CL) \\
$\beta_{\rm PN}$                                         & $1$       & $> 0.999999$ (95\% CL)      & $< 1.000004$ (95\% CL) \\
$\delta G_\mathrm{N}/G_\mathrm{N}$ (z=0)                  & $< 0.0056$ (95\% CL)       & $< 0.0041$ (95\% CL) & $< 0.0042$ (95\% CL)  \\
$10^{13} \dot{G}_\mathrm{N}/G_{\rm N}$ (z=0) [yr$^{-1}$]  & $< 0.23$ (95\% CL)        & $< 0.16$ (95\% CL)       & $< 0.19$ (95\% CL)  \\
$G_\mathrm{N}/G$ (z=0)                                    & $0.99987^{+0.00013}_{-0.000042}$  & $1.000080^{+0.000028}_{-0.000090}$          & $1.000102^{+0.000043}_{-0.00011}$  \\
\hline
$\Omega_{\rm m}$                        & $0.2999\pm 0.0050$  & $0.2994^{+0.0046}_{-0.0052}$         & $0.2997^{+0.0053}_{-0.0047}$  \\
$\sigma_8$                              & $0.8055^{+0.0059}_{-0.0069}$  & $0.8078\pm 0.0066$  & $0.8086\pm 0.0063$  \\
$S_8$                                   & $0.805\pm 0.010$           & $0.807\pm 0.010$           & $0.8082\pm 0.0094$  \\
$r_s$ [Mpc]                             & $147.63^{+0.23}_{-0.29}$   & $147.56\pm 0.25$      & $147.58^{+0.21}_{-0.24}$  \\
\hline
\end{tabular}}
\caption{\label{tab:H0} 
Constraints on the main and derived parameters (at 68\% CL if not otherwise stated) considering 
the combination with P18+BAO+SN+$p(H_0)$ for {\bf IG}, {\bf NMC+}, and {\bf NMC-}.}
\end{table*}

\begin{table*}[h!]
{\small
\centering
\begin{tabular}{l|cc}
\hline
\hline
                                         & P18 + BAO + SN & P18 + BAO + SN + $p(H_0)$  \\
\hline
$\omega_{\rm b}$                         & $0.02246\pm 0.00014$         & $0.02255\pm 0.00014$  \\
$\omega_{\rm c}$                         & $0.1194\pm 0.0010$           & $0.11900\pm 0.00099$  \\
$H_0$ [km s$^{-1}$Mpc$^{-1}$]            & $68.44^{+0.62}_{-0.79}$      & $70.18^{+0.59}_{-0.68}$  \\
$\tau$                                   & $0.0536\pm 0.0080$           & $0.0503^{+0.0085}_{-0.0073}$  \\
$\ln \left(  10^{10} A_{\rm s} \right)$  & $3.043\pm 0.016$             & $3.035^{+0.017}_{-0.015}$  \\
$n_{\rm s}$                              & $0.9671^{+0.0036}_{-0.0042}$ & $0.9687\pm 0.0038$  \\
$\xi\sigma_{\rm ini}^2$ [$M_{\rm Pl}^2$] & $> -0.0057$ (95\% CL)        & $-0.0062^{+0.0028}_{-0.0023}$  \\
$V_0$                                    & $-$                          & $< 0.81$ (95\% CL)  \\
$\sigma_{\rm ini}$ [$M_{\rm Pl}$]                &   $<0.446$ (95\% CL)          & $0.348^{+0.062}_{-0.097}$  \\
\hline
$\Omega_{\rm m}$                        & $0.3028\pm 0.0068$  & $0.2875\pm 0.0056$ \\
$\sigma_8$                              & $0.823^{+0.010}_{-0.013}$  & $0.840\pm 0.011$ \\
$S_8$                                   & $0.827\pm 0.011$           & $0.822\pm 0.010$  \\
$r_s$ [Mpc]                             & $147.00\pm 0.40$   & $146.56\pm 0.46$  \\
\hline
\end{tabular}}
\caption{\label{tab:emg} 
Constraints on the main and derived parameters (at 68\% CL if not otherwise stated) considering 
P18 in combination with BAO+SN and BAO+SN+$p(H_0)$ for the {\bf EMG} model.}
\end{table*}
\begin{table*}[h!]

        {\small
        \centering
        \begin{tabular}{l| c c c c c}
            \hline
            \hline P18+BAO+SN 
            &IG & NMC$+$ & NMC$-$ & EMG ($Z=1$) & EMG $(Z=-1)$ \\
            \hline
            {\em Planck} high-$\ell$ TTTEEE  & -1.2  & -1.3  & -1.6 & -0.8  & -0.6    \\
            {\em Planck} low-$\ell$ EE       & 0.4   & 0.4   & 1    & 0.1  & -0.3    \\
            {\em Planck} low-$\ell$ TT       & 0.2   & 0.3   & 0.3  & -0.2  & 0.5     \\
            {\em Planck} lensing             & -0.2  & -0.2  & -0.3 & 0 & 0.8     \\
            BAO                              & 0.2   & 0.2   & 0.2  & 0  & -0.5    \\
            Pantheon                         & 0.1   & 0.1   & 0.1  & 0  & -0.2     \\
            \hline
            Total                            & -0.5  & -0.5 & -0.3  & -0.9  & -0.3 \\
            \hline
    \end{tabular}}
    
    \vspace{2ex} 

        {\small
        \centering
        \begin{tabular}{l| c c c c c}
            \hline 
            P18+BAO+SN+$p(H_0)$
            &IG & NMC$+$ & NMC$-$ & EMG ($Z=1$) & EMG $(Z=-1)$ \\
            \hline
            {\em Planck} high-$\ell$ TTTEEE  & -1.2  & -0.8  & -1.4  & -0.7    & -6.5    \\
            {\em Planck} low-$\ell$ EE       & 0.8   & 0.4   & 0.7   & 0.4    & -1.8    \\
            {\em Planck} low-$\ell$ TT       & -0.1  & -0.2  & -0.2  & -0.2   & 1.1     \\
            {\em Planck} lensing             & -0.1  & -0.1  & -0.2  & -0.1      & 0.6     \\
            BAO                              & 0.1   & 0.1   & 0     & 0   & 5     \\
            Pantheon                         & 0     & 0     & 0     & 0      & 0.4     \\
            $H_0$                            & 0     & 0.2   & 0.6   & -17.8  & -13.7   \\
            \hline
            Total                            & -0.5  & -0.4  & -0.5  & -18.4  & -14.9 \\
            \hline
\end{tabular}}
\caption{\label{tab:chi_sq}
Best-fit $\Delta \chi^2$ with respect to the $\Lambda$CDM model for each dataset for the combination 
P18+BAO+SN (upper table) and P18+BAO+SN+$p(H_0)$ (lower table) for {\bf IG}, {\bf NMC+}, {\bf NMC-}, and {\bf EMG}.}
\end{table*}

\newpage

\section{Background equations} \label{sec:app_bkg}
Starting from Eq.~\ref{eqn:action}, it is possible to write down the equations governing the background 
evolution. Specializing to a spatially flat Friedmann-Lema\^itre-Robertson-Walker (FLRW) universe described by 
the line element
\begin{equation}
    \dd s^2 = a^2(\eta)\left( -\dd \eta^2 + \dd {\bf x}^2 \right)\, ,
\end{equation}
where $\eta$ is the conformal time and ${\bf x}$ the spatial comoving coordinate. The Einstein equations are obtained by varying the action \ref{eqn:action} with respect to the metric, they 
correspond to 
\begin{equation} \label{eqn:Gmunu}
    G^\mu_{\ \nu} = \frac{1}{F}\left[ T^\mu_{\ \nu} - \frac{Z}{2} \nabla^\mu \sigma \nabla_\nu \sigma 
                - g^\mu_{\ \nu}V + \left( \nabla^\mu\nabla_\nu - g^\mu_{\ \nu}\square \right) F \right]
\end{equation}
where the energy-momentum tensor for a perfect fluid is given by
\begin{equation}
    T_{\mu\nu} = p g_{\mu\nu} + (\rho + p) u_\mu u_\nu
\end{equation}
where a sum over all the species in the Universe is taken for granted, i.e. $\rho \equiv \sum_i \rho_i$ 
and $p \equiv \sum_i p_i$.

From Eq.~\eqref{eqn:Gmunu}, the Friedmann equations in Jordan frame are as follows
\begin{align} \label{eqn:Friedmann1}
    &3F\,\mathcal{H}^2 = a^2\left( \rho + V \right) + \frac{Z \sigma'^{\,2}}{2} - 3\mathcal{H} F' \\
    &-2F\,\mathcal{H}' = \frac{a^2}{3}\left(\rho + 3 p -2 V \right) + \frac{2}{3} Z \sigma'^{\,2} + F'' 
    \label{eqn:Friedmann2}
\end{align}
and the Einstein trace (the Ricci scalar) equation results
\begin{equation} \label{eqn:bkg_R}
    a^2 F R = a^2(\rho - 3p) + 4 a^2 V - 3 F_\sigma \left( \sigma'' + 2\mathcal{H}\sigma' \right)
    - (Z+3F_{\sigma\sigma})\sigma'^{\,2} \,.
\end{equation}
Finally, the evolution equation of the scalar field $\sigma$ is governed by the modified Klein-Gordon 
equation
\begin{align}
    \sigma'' + 2\mathcal{H}\sigma' - \frac{F_\sigma}{2ZF + 3F_\sigma^2}
    \Bigl[&a^2(\rho - 3p) + 4 a^2 \left(V - \frac{F}{2F_\sigma}V_\sigma\right) \notag\\
    & - \left(Z + 3F_{\sigma\sigma}\right)\sigma'^{\,2}\Bigr] = 0 \,.
\end{align}

\section{Linear perturbed equations} \label{sec:app_pert}
In the synchronous gauge, up to linear order in perturbed quantities, the perturbed FLRW metric is
\begin{equation} \label{eqn:pert_metric}
    \dd s^2 = a^2(\eta) \left[-\dd^2 \eta + (\delta_{ij} + h_{ij}) \dd x^i \dd x^j \right]
\end{equation}
where $h_{ij}$ is the metric perturbation.

From this point on, we move in Fourier space for the calculation of the perturbed quantities.
The scalar mode of $h_{ij}$ can be express as a Fourier integral as
\begin{equation}
    h_{ij}(\eta,\,{\bf x}) = \int \dd^3k e^{\imath {\bf k\cdot x}}
                            \left[\hat{k}_i\hat{k}_jh(\eta,\,{\bf k})
                            + \left(\hat{k}_i\hat{k}_j-\frac{\delta_{ij}}{3}\right)6\xi(\eta,\,{\bf k})\right]
\end{equation}
where $\hat{k}_i = k_i/k$ with $k = |{\bf k}|$ and $h\equiv \delta^{ij} h_{ij}$ is the Fourier 
transform of trace of $h_{ij}(\eta,\,{\bf x})$. 
We follow the conventions of Ref.~\cite{Ma:1995ey}.

\subsection{The perturbed Einstein field equations}
Splitting the Einstein tensor as the sum of the background (mean) part $\bar{G}_{\mu\nu}$ and its 
corresponding perturbation $\delta G_{\mu\nu}$, i.e. $G_{\mu\nu} = \bar{G}_{\mu\nu}+\delta G_{\mu\nu}$, 
the scalar perturbations in synchronous gauge are usually presented as time-time, longitudinal 
time-space, trace space-space, and longitudinal traceless space-space parts of the Einstein equations 
in Fourier space as follow
\begin{align}
    \xi k^2 - \frac{{\cal H}h'}{2} &= a^2 \frac{\delta G^0_{\ 0}}{2} \notag\\
                                   &= 4\pi G a^2 \delta T^0_{\ 0} \,,\\
    k^2 \xi' &= a^2 \frac{\nabla^i \delta G^0_{\ i}}{2} \notag\\
             &= 4\pi G a^2 (\bar{\rho}+\bar{p})\theta \,,\\
    h'' + 2{\cal H}h' - 2\xi k^2 &= -a^2 \delta G^i_{\ i} \notag\\
             &= -8\pi G a^2 \delta T^i_{\ i} \,,
\end{align}
\begin{align}
    h'' + 6\xi'' + 2{\cal H}(h' + 6\xi') - 2k^2\xi &= 
            -3 a^2 \left(\hat{k}_i\hat{k}_j-\frac{\delta_{ij}}{3}\right) \delta G^i_{\ i} \notag\\
            &= -24\pi G a^2 (\bar{\rho}+\bar{p})\Theta
\end{align}
where we used the definition
\begin{align}
    &(\bar{\rho}+\bar{p})\theta \equiv \imath k^i \delta T^0_{\ i} \,,\\
    &(\bar{\rho}+\bar{p})\Theta \equiv -\left(\hat{k}_i\hat{k}_j-\frac{\delta_{ij}}{3}\right) \Sigma^i_{\ j} \,,\\
    &\Sigma^i_{\ j} \equiv \bar{T}^i_{\ j} - \delta^i_j \frac{\bar{T}^k_{\ k}}{3} \,.
\end{align}
and splitting the total energy density and pressure in a background and perturbed parts, we obtain 
the following elements
\begin{align} 
    &\bar{T}_{00} = a^2(\bar{\rho} + \delta\rho) \,,\\
    &\bar{T}_{0i} = -a^2(\bar{\rho} + \bar{p})v^i \,,\\ 
    &\bar{T}_{ij} = a^2\delta_{ij}\left(\bar{p} + \delta p\right) + a^2\Sigma_{ij} \,.
\end{align}

The first perturbed equation involving total density fluctuations reads
\begin{equation}
    \xi k^2 - \frac{{\cal H}h'}{2} = -(8\pi G) a^2 \frac{\delta \tilde{\rho}}{2 F}
\end{equation}
with
\begin{align}    
    \delta \tilde{\rho} = &\delta \rho - \frac{h'\bar{\sigma}'\bar{F}_\sigma}{2a^2}
        +\frac{\delta\sigma'}{a^2} \left(Z\bar{\sigma}'-3{\cal H}\bar{F}_\sigma\right)         -\frac{\delta\sigma\bar{F}_\sigma}{a^2\bar{F}} \left[a^2\bar{\rho} + \frac{Z}{2}\bar{\sigma}'^2
        \right. \notag\\
        &\left.+a^2\left(V - V_\sigma\frac{F}{F_\sigma}\right)-3{\cal H}\bar{F}_\sigma\bar{\sigma}' + 3{\cal H}\frac{F_{\sigma\sigma}F}{F_\sigma}\bar{\sigma}' 
        + k^2F\right] \,.
\end{align}

The second perturbed equation involving total velocity reads
\begin{equation}
    k^2 \xi' = (8\pi G) a^2 \frac{(\tilde{\rho}+\tilde{p})\tilde{\theta}}{2 F}
\end{equation}
with
\begin{equation}
    \left(\tilde{\rho}+\tilde{p}\right)\tilde{\theta} = 
        \left(\bar{\rho}+\bar{p}\right)\theta + \frac{k^2}{a^2}
        \left[\left(Z\bar{\sigma}'-{\cal H}F_\sigma+F_{\sigma\sigma}\bar{\sigma}'\right)\delta\sigma
        +F_\sigma\delta\sigma'\right] \,.
\end{equation}

The third perturbed equation involving total pressure reads
\begin{equation}
    h'' + 2{\cal H}h' - 2k^2\xi = -3 (8\pi G) a^2 \frac{\delta\tilde{p}}{F}
\end{equation}
with
\begin{align}
    \delta\tilde{p} = &\delta p + \frac{h'F'}{3a^2} + \frac{Z}{a^2}\bar{\sigma}'\delta\sigma' - \delta V +     \frac{2}{3a^2}k^2 \delta F 
        + \frac{{\cal H}}{a^2}\delta F' + \frac{\delta F''}{a^2} \notag\\
        &-\delta\sigma\frac{\bar{F}_\sigma}{a^2\bar{F}}\left(a^2\bar{p} + \frac{Z}{2}\bar{\sigma}'^2 - a^2\bar{V} + {\cal H}F' + F''\right) \,.
\end{align}

The fourth perturbed equation involving total shear reads
\begin{equation}
        h'' + 6\xi'' + 2{\cal H}(h' + 6\xi') - 2k^2\xi = -3(8\pi G) a^2 \frac{(\tilde{\rho}+\tilde{p})\tilde{\Theta}}{F}
\end{equation}
with
\begin{equation}
        (\tilde{\rho}+\tilde{p})\tilde{\Theta} = (\bar{\rho}+\bar{p})\Theta 
        + \frac{2 k^2}{3a^2} \left(F_\sigma \delta\sigma + F'\frac{h'+6\xi'}{2k^2}\right) \,.
\end{equation}

The perturbed Ricci scalar is given by
\begin{align}
     a^2F \delta R =\, &a^2(\delta\bar{\rho} - 3\delta\bar{p}) - \frac{3h'F'}{2}
        -2Z\bar{\sigma}'\delta\sigma' + 4a^2V_\sigma\delta\sigma - 6{\cal H}\delta F' - 3\delta F''  \notag\\
        &- 3k^2\delta F - \frac{\delta\sigma F_\sigma}{F} 
        \left[a^2(\bar{\rho} - 3\bar{p}) - Z\bar{\sigma}'^2 + 4a^2\bar{V} - 3F'' - 6{\cal H}F'\right] \,.
\end{align}

\subsection{The perturbed Klein-Gordon equation}
The perturbed equation for the evolution of the scalar field perturbation $\delta\sigma$ is
\begin{equation}
    Z\delta\sigma''+2{\cal H}Z\delta\sigma'+\left[Zk^2+a^2\left(V_{\sigma\sigma}-\frac{RF_{\sigma\sigma}}{2}\right)\right]\delta\sigma 
    + Z\frac{h'\bar{\sigma}'}{2} - \frac{a^2F_\sigma}{2}\delta R = 0 \,.
\end{equation}

\section{Initial conditions} \label{sec:app_IC}
The adiabatic initial condition for the background correspond to
\begin{align}
	a(\tau) &= \sqrt{\frac{\rho_{\rm r\,0}}{3F_{\rm ini}}}\tau\left[1 + \frac{Z}{4}\omega\tau  
	- \frac{5ZF_{\rm ini,\sigma}^2\left(Z+3F_{\rm ini,\sigma\sigma}\right)}{64ZF_{\rm ini}+96F_{\rm ini,\sigma}^2}\left(\omega\tau\right)^2\right] \,,\notag\\
	\mathcal{H}(\tau) &= \frac{1}{\tau}\left[1+\frac{Z}{4}\omega\tau 
	- Z \frac{2F_{\rm ini} + F_{\rm ini,\sigma}^2\left(8Z+15F_{\rm ini,\sigma\sigma}\right)}{32ZF_{\rm ini} + 48F_{\rm ini,\sigma}^2}\left(\omega\tau\right)^2\right] \,,\notag\\
	\sigma(\tau) &= \sigma_{\rm ini} + \frac{3F_{\rm ini,\sigma}}{4}\omega\tau - F_{\rm ini,\sigma}\frac{4Z F_{\rm ini}(2Z - 3F_{\rm ini,\sigma\sigma}) + 27F_{\rm ini,\sigma}^2\left(Z+F_{\rm ini,\sigma\sigma}\right)}{32( 2 Z F_{\rm ini}+3F_{\rm ini,\sigma}^2)}\left(\omega\tau\right)^2
\end{align}
where 
\begin{equation}
    \omega = \frac{\rho_{\rm m,\,0}}{\sqrt{3\rho_{\rm r,\,0}}}\frac{2\sqrt{F_{\rm ini}}}{2Z F_{\rm ini} + 3F_{\rm ini,\sigma}^2} \,.
\end{equation}

For the cosmological fluctuations in the synchronous gauge, we have as adiabatic initial conditions
\begin{align}
    &\delta_\gamma(k\,,\tau) = \delta_\nu(k\,,\tau) = \frac{4}{3}\delta_{\rm b}(k\,,\tau) = \frac{4}{3} \delta_{\rm c}(k\,,\tau) = -\frac{(k\tau)^2}{3}\left(1-\frac{Z\omega\tau}{5}\right)\\
    &\theta_\nu(k\,,\tau) = -\frac{k^4\tau^3}{36}\frac{23 + 4R_\nu}{15 + 4R_\nu}
    \left[1 - \frac{3}{20}\frac{Z\left(275 + 50 R_\nu + 8R_\nu^2\right)F_{\rm ini} + 15(5 - 4R_\nu)F_{\rm ini,\sigma}^2}{(15 + 2R_\nu)(23 + 4R_\nu)F_{\rm ini}}\omega\tau\right] \\
    &\theta_\gamma(k\,,\tau) = \theta_{\rm b}(k\,,\tau) = -\frac{k^4\tau^3}{36}\left[1 - \frac{3}{20} \frac{Z(1-R_\nu+5R_{\rm b})F_{\rm ini} + \frac{15}{2}R_{\rm b}F_{\rm ini,\sigma}^2}{\left(1 - R_\nu\right)F_{\rm ini}}\omega\tau\right]
    \\
    &\theta_{\rm c}(k\,,\tau) = 0 \\
    &\sigma_\nu(k\,,\tau) = \frac{2 (k \tau)^2}{3 (15 + 4R_\nu)} \left[1 + 
    \frac{(-5+4R_\nu)\left(2ZF_{\rm ini} + 3F_{\rm ini,\sigma}^2\right)}{8(15+2R_\nu)F_{\rm ini}}\omega\tau\right]\\
    &h(k\,,\tau) = \frac{\left(k\tau\right)^2}{2} \left(1 - \frac{Z \omega \tau}{5}\right) \\
    &\eta(k\,,\tau) = 1 - \frac{\left(k\tau\right)^2}{12(15+4R_\nu)} \left[5 + 4R_\nu
    -\frac{2Z(5+ 4R_\nu) (65 + 4R_\nu)F_{\rm ini} + 75(-5 + 4R_\nu)F_{\rm ini,\sigma}^2}{20 (15 + 2R_\nu)F_{\rm ini}}\omega\tau\right] \\
    &\delta\sigma(k\,,\tau) = -\frac{1}{16}F_{\rm ini,\sigma} k^2 \tau^3\omega \left[1 - \frac{2Z(8Z - 9F_{\rm ini,\sigma\sigma})F_{\rm ini} + (45F_{\rm ini,\sigma\sigma} + 48Z)F_{\rm ini,\sigma}^2}{40ZF_{\rm ini} + 60F_{\rm ini,\sigma}^2}\omega\tau\right] 
\end{align}
where $R_\nu = \rho_{\nu,\,0}/\rho_{\rm r,\,0}$ and $R_{\rm b} = \rho_{\rm b,\,0}/\rho_{\rm m,\,0}$.
These quantities reduce to induced gravity for $Z = 1$ and $F = \xi\sigma^2$ \cite{Paoletti:2018xet}, 
to a non-minimally coupled scalar field with standard kinetic term for $Z = 1$ and 
$F = N_{\rm Pl}^2+\xi\sigma^2$ \cite{Rossi:2019lgt}, and General Relativity for $Z = 1$ and 
$F = M_{\rm Pl}^2$ \cite{Ma:1995ey}.

\section{Comparison between BAO, and FS + BAO joint analysis} \label{sec:app_bao}
We compare here the results adding different datasets of galaxy information to the {\em Planck} DR3 data 
such as the full shape (FS) of BOSS DR12 pre-reconstructed power spectrum measurements 
\cite{Gil-Marin:2015sqa,DAmico:2020kxu}, BAO of BOSS DR12 post-reconstruction power spectrum measurements \cite{BOSS:2016wmc}, low-$z$ BAO measurements from 
SDSS DR7 6dF and MGS \cite{Beutler:2011hx,Ross:2014qpa}, Ly$\alpha$ BAO measurements from eBOSS 
\cite{deSainteAgathe:2019voe,Blomqvist:2019rah,Cuceu:2019for}, and combination of those including the 
covariance among the DR12 datasets.

\begin{figure}
\centering
\includegraphics[width=0.98\textwidth]{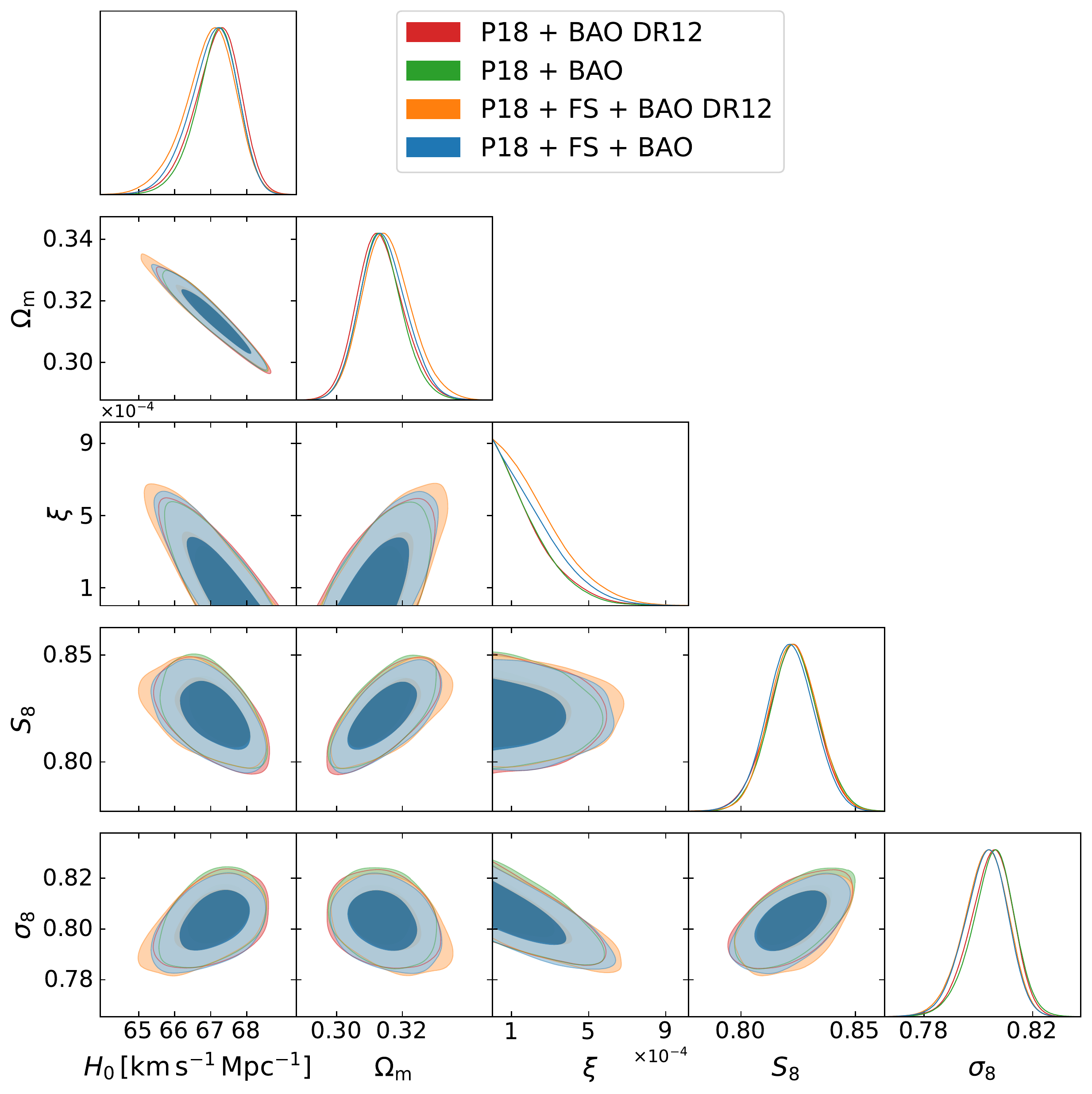}
\caption{Marginalized joint 68\% and 95\% CL regions 2D parameter space using the CMB P18 data for IG 
($F = \xi\sigma^2$) in the phantom branch ($Z = -1$) in combination with BAO from BOSS DR12 (red), 
BAO from BOSS DR12/SDSS DR7 6dF-MGS/eBOSS (green), FS plus BAO from BOSS DR12 (orange), and FS plus 
BAO from BOSS DR12/SDSS DR7 6dF-MGS/eBOSS (blue).}
\label{fig:ig_ph_bao}
\end{figure}
In Fig.~\ref{fig:ig_ph_bao}, we show the marginalized posterior distributions of the cosmological 
parameters for IG with P18 plus different combinations of the FS and BAO measurements. 
We see that the posterior distributions are very robust among the combination considered and that 
the addition of FS information to the combination P18+BAO does not change the marginalized constraints 
for the models studied here.

\newpage
\bibliographystyle{unsrtnat}
\bibliography{Biblio}

\end{document}